\RequirePackage[colorlinks,allcolors=blue,hypertexnames=false]{hyperref}
\RequirePackage[dvipsnames]{xcolor}
\documentclass[reprint]{agujournal2019}

\usepackage{amsmath, amssymb, amsthm}
\usepackage{mathtools}
\usepackage{gensymb}
\usepackage{bm}
\usepackage{caption}
\usepackage{ragged2e}
\usepackage{xcolor}

\usepackage{graphicx}

\usepackage{url} 
\usepackage{soul}

\usepackage{float}
\usepackage{algorithm}
\usepackage{algpseudocode}
\usepackage{matlab-prettifier}
\usepackage{tabularx}

\theoremstyle{definition}

\newcommand{\vc}{\mathbf}

\journalname{JGR: Machine Learning and Computation}

\begin{document}
\justifying

\title{Rapid estimation of global sea surface temperatures from sparse streaming in situ observations\thanks{Accepted for publication in the Journal of Geophysical Research: Machine Learning and Computation}}
\authors{Cassidy All\affil{1}, Kevin Ho\affil{2}, Maya Magnuski\affil{3}, Christopher Nicolaides\affil{4}, Louisa B. Ebby\affil{5}, Mohammad Farazmand\affil{5}} 
\affiliation{1}{Department of Applied Mathematics, University of Colorado Boulder, 11 Engineering Drive, Boulder, CO 80309}
\affiliation{2}{Department of Mathematics and Statistics, Mississippi State University, 175 President's Circle, MS 39762}
\affiliation{3}{Mathematics Program, Bard College, 30 Campus Road, Annandale-On-Hudson, NY 12504}
\affiliation{4}{Department of Mathematics, Indiana University, 831 East 3rd Street, Bloomington, IN 47405}
\affiliation{5}{Department of Mathematics, North Carolina State University, 2311 Stinson Drive, Raleigh, NC 27695}

\correspondingauthor{M. Farazmand}{farazmand@ncsu.edu}

\begin{keypoints}
	\item We introduce S-DEIM for accurate reconstruction of spatiotemporal geophysical quantities from extremely sparse observations.
	\item S-DEIM reconstructions are achieved through a synergistic and rigorous combination of empirical interpolation and deep learning.
	\item By applying it to SST reconstruction, we demonstrate that S-DEIM is robust to observational noise and sensor placement method.
\end{keypoints}

\begin{abstract}
Reconstructing high-resolution sea surface temperatures (SST) from staggered SST measurements is essential for analyzing earth system processes. However, when SST measurements are sparse, the resulting inferred SST fields are rather inaccurate. Here, we show that Sparse Discrete Empirical Interpolation Method (S-DEIM) can be used as a model-free data assimilation method to reconstruct high-resolution SST fields from sparse in situ observations. The S-DEIM estimate consists of two terms, one computed from instantaneous in situ observations using empirical interpolation, and the other learned from the historical time series of observations using recurrent neural networks (RNNs). We train the RNNs using the National Oceanic and Atmospheric Administration's weekly high-resolution SST dataset spanning the years 1989-2021 which constitutes the training data. Subsequently, we examine the performance of S-DEIM on the test data, comprising January 2022 to January 2023. For this test data, S-DEIM infers the high-resolution SST from 100 in situ observations, constituting only 0.2\% of the high-resolution spatial grid. We show that the resulting S-DEIM reconstructions are about 40\% more accurate than earlier empirical interpolation methods, such as DEIM and Q-DEIM. Furthermore, 91\% of S-DEIM estimates fall within $\pm 1^\circ$C of the true SST. We also demonstrate that S-DEIM is robust with respect to sensor placement: even when the sensors are distributed randomly, S-DEIM reconstruction error deteriorates only by 1-2\%. S-DEIM is also computationally efficient: training the RNN, which is performed only once offline, takes approximately one minute. Once trained, the S-DEIM reconstructions are computed in less than a second. 
\end{abstract}

\section*{Plain Language Summary}
Monitoring, forecasting, and analyzing ocean-atmosphere processes rely on high-resolution geophysical quantities, such as temperature, wind, and humidity. When observational data is sparse, estimating these spatiotemporal quantities becomes imprecise. In this study, we investigate the ability of the Sparse Discrete Empirical Interpolation Method (S-DEIM)---a data-driven, model-free method---to infer high-resolution quantities from limited observations. S-DEIM integrates conventional empirical interpolation with modern deep learning techniques through a mathematically rigorous framework. We demonstrate its efficacy by reconstructing global sea surface temperatures, achieving results in seconds with a 40\% improvement in accuracy over conventional interpolation methods.

\section{Introduction}
Sea surface temperature (SST) is a critical piece of information for understanding various meteorological and oceanographic processes. For example, short-term weather forecasts~\cite{chelton2005}, seasonal to long-term climate predictions~\cite{banzon2016,goddard2001}, and understanding marine ecosystems~\cite{Venegas2023} rely on sea surface temperature measurements. 

SST observations are gathered by two predominant modalities: remote sensing through satellites and direct in situ observations through buoys, drifters, and ships. In situ measurements provide highly accurate local SST measurements but are limited in number due to logistical and economic constraints. Although satellite-based measurements via infrared and microwave radiometers offer broader spatial coverage, cloud cover and atmospheric conditions can interfere with the accuracy of their observations~\cite{Koner2016}. Therefore, SST measurements alone do not provide accurate and high-fidelity SST fields. Instead, as we review in Section~\ref{sec:relWork}, these measurements are used in a data assimilation process to obtain accurate and high-resolution estimates of the SST field.

The main objective of the present paper is to investigate the efficacy of a new data assimilation method, namely the Sparse Discrete Empirical Interpolation Method (S-DEIM) \cite{Farazmand2024}, for estimating high-resolution SST fields from very sparse in situ observations. S-DEIM consists of two components. One component is determined from instantaneous in situ observations, using classical interpolation techniques. The other component is estimated from time series of past in situ observations, using recurrent neural networks (RNNs)~\cite{Farazmand2024b}.

The resulting RNN-based S-DEIM method is model-free, i.e., it does not require a geophysical fluid dynamics model. Furthermore, it reproduces reasonably accurate SST estimates even when in situ observations are very sparse. Finally, the estimates can be reproduced rapidly, within a few seconds, rendering it particularly attractive for real-time nowcasting of SST fields. The S-DEIM estimates are not as accurate as model-assisted data assimilation methods, such as variational methods and ensemble Kalman filtering (see Section~\ref{sec:relWork} for a review). However, the S-DEIM estimate provides an excellent initial guess for such model-assisted data assimilation techniques.

\subsection{Background and Related Work}\label{sec:relWork}
In this section, we review different types of available observational data and data assimilation methods that are common in physical oceanography. We also discuss previous studies that use DEIM-based methods in the context of SST reconstruction. 

 \emph{Observational data:} SST observations are collected primarily through two methods: (i) remote sensing and (ii) direct in situ measurements~\cite{Kent2010}. Remote sensing is performed through satellites that carry infrared and microwave sensors~\cite{Dickey2006}. Although satellites offer broad spatial coverage, they measure radiance, which can be affected by atmospheric conditions, leading to high measurement uncertainty. Additionally, infrared sensors cannot make observations through clouds, contributing to gaps in satellite coverage~\cite{Minnett2019}. On the other hand, in situ measurements, collected directly using moored buoys, ships, and surface drifters, tend to be more accurate~\cite{Kennedy2013}. However, these in situ observations are limited by their relatively sparse spatial distribution. As we show here, S-DEIM is capable of producing reasonably accurate SST reconstructions from these sparse observations. As we discuss in Section~\ref{subsec:cpqr}, S-DEIM assumes static sensors whose positions do not change in time. Therefore, our method, in its current form, is limited to using in situ observations from moored buoys.

 \emph{Reanalysis:} In oceanography, the term ``reanalysis" refers to the blending of high-fidelity geophysical models and observational measurements through data assimilation to improve the accuracy of SST estimation and fill the gaps where observational data is lacking. The choice of the data assimilation method must strike a balance between computational cost and the desired accuracy. For example, variational data assimilation methods~\cite{Courtier1998,Courtier1994,cummings2013} are the most accurate, but demand substantial computational resources, since they require several forward solves of the geophysical model and backward-in-time solves of the model adjoint. A computationally less demanding method is the ensemble Kalman filter, which requires repeated solves of the forward model, but does not require the adjoint equations~\cite{Evensen2003,Rozier2007,Gorthi2011}. The National Oceanic and Atmospheric Administration (NOAA) uses Optimum Interpolation (OI) to create high-resolution SST fields, producing weekly and daily datasets~\cite{Reynolds2002}. OI is similar to Kalman filtering,  but it simplifies the data assimilation process by assuming a static covariance matrix to reduce computational cost~\cite{1994_OI, Reynolds2007}. In contrast to reanalysis, the proposed S-DEIM method does not require a geophysical model for its reconstructions.

\emph{Model-free SST reconstruction:} In the past decade, with the advent of deep learning and artificial intelligence, there has been a growing interest in purely data-driven reconstruction and forecast of geophysical quantities, such as SST~\cite{Bi2023,Bodnar2025,pathak2022}. These model-free methods use the abundant reanalysis data, going back several decades, to train deep learning models. Once trained, the deep learning model is in turn used for rapid SST reconstruction and forecast~\cite{Taylor2022,Xu2023}. A promising deep learning tool is the so-called autoencoders which first learn a mapping from high-fidelity reanalysis data to sparse observations. Then a second mapping is learned from sparse observations back to the high-fidelity data. The two maps are constrained so that their composition approximates the identity mapping. Although autoencoders have had some success in reconstruction and forecast of regional SST~\cite{Barth2022,Goh2024,Lobashev2023}, they are not yet adept at reconstructing global SST from sparse observations.

\emph{DEIM and Q-DEIM:} DEIM is a particular example of a model-free method for SST reconstruction, which is notable for its interpretability and low computational cost. Although originally developed for reduced-order modeling~\cite{Barrault2004,Sorensen2010}, it was later repurposed for flow reconstruction from incomplete observations~\cite{Manohar2018}. As an empirical interpolation, DEIM partially owes its success to the empirical basis functions (or modes) in which the interpolant is expanded. Instead of using a prescribed basis, such as polynomials, DEIM uses an empirical basis which is derived specifically for the system of interest. In the case of SST reconstructions, the basis is derived by performing Proper Orthogonal Decomposition (POD) on historical reanalysis data. In this regard, DEIM is similar to the Empirical Orthogonal Functions (EOFs) method~\cite{Beckers2003,Smith1996}. The expansion coefficients within the POD basis are obtained by solving a linear least squares problem (see Section~\ref{sec:sdeim}).

The quality of DEIM reconstructions is highly sensitive to the locations of the sensor measurements. \citeA{Drmac2016} introduced the Q-DEIM method, which uses column-pivoted QR (CPQR) factorization to approximate optimal sensor locations. Alternative sensor placement methods in the context of SST reconstruction have also been explored. For example, \citeA{Saito2021} developed a determinant-based sensor placement method that seeks to minimize uncertainties in the reconstruction. \citeA{Brunton2021} consider the multi-fidelity sensor placement problem with cost constraints, introducing greedy algorithms for placing a mix of sensors with varying accuracy. \citeA{Klishin2023} introduce a sensor placement method based on ideas from statistical physics. We refer to \citeA{karnik2025} for a comprehensive review of sensor placement methods.

Another aspect of DEIM which has received much attention is its sensitivity to observational noise~\cite{argaud2017,Drmac2020}. In particular, \citeA{Callaham2019} investigate a variety of sparsity-promoting optimization problems instead of linear least squares. They find that sparse representations improve the robustness of the reconstructions to observational noise.

\emph{S-DEIM:} DEIM was originally developed under the assumption that the number of sensors $r$ is equal to the number of modes $m$~\cite{Sorensen2010}. The overdetermined regime, where $r>m$, has also received much attention. In fact, when $r>m$, DEIM coincides with the gappy POD method~\cite{Sirovich1995}, which predates it by more than a decade. Numerical evidence suggests that DEIM estimates become more accurate as the number of sensor measurements increases for a fixed number of modes~\cite{Brunton2021}. In the overdetermined regime ($r > m$), there is also evidence that the reconstructions are more robust to observational noise~\cite{argaud2017,Drmac2020}. 

In contrast, the underdetermined regime ($r<m$), where fewer sensor measurements are available, has only recently attracted attention~\cite{Farazmand2024}. This regime can occur when the number of sensor measurements $r$ is very limited, but $m=r$ modes do not provide enough model complexity to faithfully reconstruct the desired field. \citeA{Farazmand2024} introduced S-DEIM, extending DEIM to the underdetermined case ($r < m$).
S-DEIM adds the previously ignored kernel vector to the solution of the least squares problem underlying DEIM.
Unfortunately, the optimal value of this kernel cannot be evaluated from sensor measurements alone. 
\citeA{Farazmand2024} introduced a data assimilation method that leverages the governing equations of the system to estimate this unknown kernel vector. More recently, it was also demonstrated that this kernel vector can be estimated as the output of a recurrent neural network whose inputs comprise the time series of the observational data, including past history~\cite{Farazmand2024b}. As a result, S-DEIM can be used as a model-free data assimilation method. This RNN-based S-DEIM has proven extremely effective on synthetic data, i.e., data produced as numerical solutions of differential equations~\cite{Farazmand2024b}. The present paper is the first time that its efficacy is being tested on truly empirical data, i.e., NOAA's sea surface temperature data.

\subsection{Outline}
The structure of the paper is as follows. In Section~\ref{sec:data}, we describe the datasets used in this paper. Section~\ref{sec:sdeim} reviews S-DEIM and outlines the sensor placement methods. In Section~\ref{sec:recurrent}, we discuss estimation of the S-DEIM kernel vector using recurrent neural networks. We present the numerical results in Section~\ref{sec:results}. Finally, Section~\ref{sec:conclusion} concludes our findings and presents directions for future work. 

\section{Datasets Used}\label{sec:data} 
\begin{figure*}[h]
	\centering
	\includegraphics[width=\textwidth]{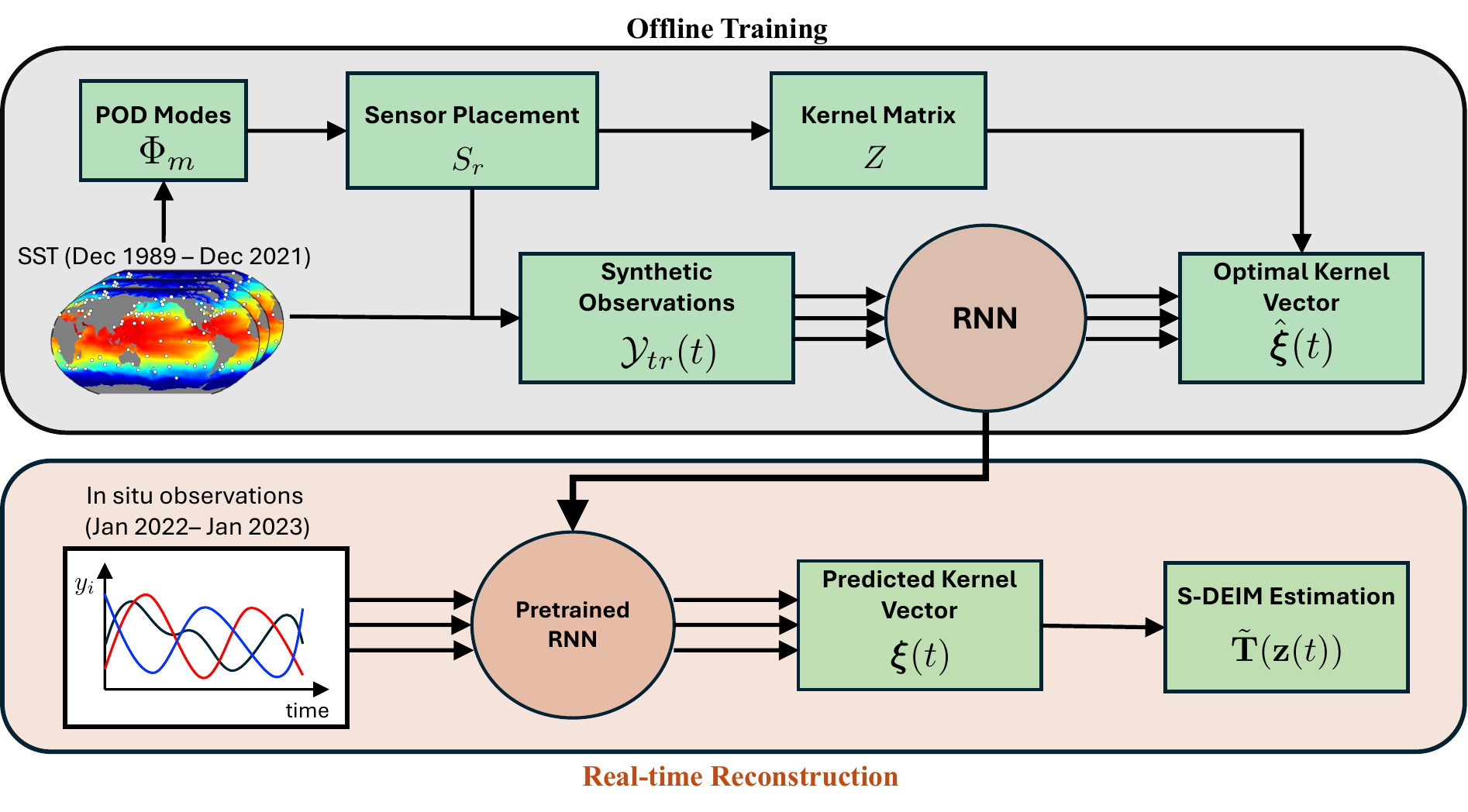}
	\caption{Schematic framework for S-DEIM. In the offline stage, we use high-fidelity data to train the RNN. This pretrained RNN is then used in real-time reconstruction to obtain the S-DEIM estimate from sparse in situ observations.}
	\label{fig:my_pipeline}
\end{figure*}

The proposed framework is depicted in Figure~\ref{fig:my_pipeline}. Each compartment of this framework is elaborated in Sections~\ref{sec:sdeim} and~\ref{sec:recurrent}. It consists of two primary components: the offline training phase, which requires high-fidelity SST data, and a real-time reconstruction phase which only relies on sparse in situ observations.

Throughout this paper, we use the weekly NOAA Optimal Interpolation SST (OISST) dataset, spanning the period from December 1989 to January 2023~\cite{Reynolds2002}. We split this dataset into two subsets: 
\begin{enumerate}
    \item Training data (December 1989 - December 2021): This comprises our high-fidelity data for the offline training phase. The weekly SST is stored on a spatial grid of $360\times 180$, i.e., one grid point per degree in both longitudinal and latitudinal directions.
    \item Test data (January 2022 - January 2023): This data is used to extract observational time series at the sensor locations. Only these sparse time series, not the high-fidelity data, are used to obtain the S-DEIM reconstructions. Therefore, the test data should be interpreted as a proxy for in situ observations when high-fidelity SST data is unavailable.
    The high-fidelity test data are only used on the backend to compute and report reconstruction errors. The test data corresponds to a weak to moderate La Ni\~na year with an Oceanic Ni\~no Index (ONI) value between $-0.8\,^{\circ}\text{C}$ and $-1.1\,^{\circ}\text{C}$ \cite{diamond2023}.
\end{enumerate}

In oceanography, it is customary to remove the temporal average of SST and work with SST anomalies~\cite{Deser2010}. Here, we follow this convention by computing the SST time-average of the training data and removing it from both training and test data. Centering the data through this procedure also helps improve the quality of the extracted POD modes~\cite{berkooz93}. The temperature can be easily recovered by adding the training mean back to the reconstructed SST anomalies. In the following sections, we simply say SST instead of SST anomaly, unless otherwise stated explicitly.

\section{Methods}
\label{sec:sdeim}
Consider the sea surface temperature anomalies $T(\vc x,t)$, where $t$ denotes time and $\vc x\in\Omega$ is a point on the surface of the globe $\Omega$. The NOAA dataset is discretized over the spatial grid $\mathcal G$ which contains $360\times 180$ grid points, one grid point per degree. We denote the grid points with $\mathcal{G} := \{\mathbf{x}_i\}_{i=1}^N \subset \Omega$ where $N=44,219$, excluding the grid points that fall on land.
At each time $t$, we store the SST as a vector,
\begin{equation}
    \mathbf{T}(t) := \left[
        T(\mathbf{x}_1,  t), T(\mathbf{x}_2, t),  \ldots, T(\mathbf{x}_N, t)
    \right]^\top \in \mathbb{R}^N.
\end{equation}

Our goal is to infer the global SST $\mathbf{T}(t)$ from in situ observations. More precisely, we assume only $r$ entries of $\mathbf{T}(t)$ are known through direct observations, where $r\ll N$. In practice, the known measurements are obtained from $r$ sensors, e.g., buoys. Let $\{i_k\}_{k=1}^r \subset \{1, 2, \ldots, N\}$ denote the indices of the sensor locations. We collect the observations in the vector $\mathbf{y}(t) := \left[ \mathbf{T}_{i_1}(t),  \ldots, \mathbf{T}_{i_r}(t)\right]^\top + \bm{\epsilon}$ where $\bm{\epsilon} \in \mathbb{R}^r$ represents observational noise. For simplicity, we neglect observational noise in this section and set $\bm{\epsilon} = \vc 0$.

We encode the location of $r$ sensors in a \textit{selection matrix} $S_r := \left[\mathbf{e}_{i_1} \vert \mathbf{e}_{i_2} \vert \ldots \vert \mathbf{e}_{i_r} \right]$, where $\{\mathbf{e}_{i}\}_{i=1}^N$ denote the standard Euclidean bases in $\mathbb R^N$. Then, the observations at time $t$ can be expressed as
\begin{equation}
\mathbf{y}(t) = S_r^\top \mathbf{T}(t).
\end{equation}

We seek to estimate the complete state $\mathbf{T}(t)$ from these observations $\mathbf{y}(t)$. For notational simplicity, we may omit the dependence of variables on time $t$.
We estimate the vector $\mathbf{T}$ within an orthonormal set $\{\bm{\phi}_1,\ldots,\bm{\phi}_m\} \subset \mathbb{R}^N$ so that 
\begin{equation}
\mathbf{T}(t) \simeq \sum_{i=1}^m \alpha_i(t)\bm{\phi}_i = \Phi_m \bm{\alpha}(t).
\end{equation}
Here, $\Phi_m := \left[\bm{\phi}_{1} \vert \bm{\phi}_{2} \vert \ldots \vert \bm{\phi}_{m} \right] \in \mathbb{R}^{N \times m}$ is the \textit{basis matrix} with $m$ basis vectors, and the coefficients $\bm{\alpha} = \left[\alpha_1, \alpha_2, \ldots, \alpha_m\right]^\top \in \mathbb{R}^m$ represent coordinates in the basis $\{\bm\phi_i\}_{i=1}^m$. As we discuss in Section~\ref{subsec:basis}, the basis matrix is usually derived from POD analysis of the training dataset.
As the basis matrix $\Phi_m$ is truncated to include only $m$ modes, any reconstruction $\Phi_m\bm{\alpha}$ incurs some truncation error, 
\begin{equation}
    \mathcal{E}_m(\bm{\alpha}):= \|\Phi_m \bm{\alpha} - \mathbf{T}\|,
\end{equation}
where $\|\cdot\|$ denotes the standard Euclidean norm. Consider the coefficients $\bm{\alpha}$ that minimize the truncation error, 
\begin{equation}
\label{eq:og_min_problem}
\hat{\pmb\alpha}:=\arg\min_{\bm{\alpha} \in \mathbb{R}^m}
\mathcal{E}_m(\bm{\alpha}), 
\end{equation}
 where the unique minimizer is given by $\hat{\bm{\alpha}}=\Phi_m^\top\mathbf{T}$. The coefficients $\hat{\bm{\alpha}}$ determine the \textit{best fit approximation},
\begin{equation}
\label{eq:best_fit_approx}
\hat{\mathbf{T}}:=\Phi_m \hat{\bm{\alpha}}=\Phi_m\Phi_m^\top\vc T,
\end{equation}
which represents the orthogonal projection of $\mathbf{T}$ onto the subspace spanned by the columns of the basis matrix $\Phi_m$. The truncation error $\mathcal E_m(\hat{\bm\alpha})$ decreases monotonically as the number of modes $m$ increases. 

\subsection{S-DEIM}
Unfortunately, the best fit approximation \eqref{eq:best_fit_approx} is not directly computable because it requires knowledge of the complete SST $\mathbf{T}$. Instead, we define the observation error, 
\begin{equation}
\label{eq:deim_err_term}
 \mathcal{E}_{obs}(\bm{\alpha}) := \left\|S_r^\top \Phi_m \bm{\alpha} - \mathbf{y} \right\|.
\end{equation}
Note that $S_r^\top \Phi_m \bm{\alpha}$ corresponds to the entries of the reconstruction $\Phi_m\bm\alpha$ at the sensor locations. Since the observations $\mathbf{y}$ are known, we can compute the observation error $\mathcal{E}_{obs}$ for a given $\bm\alpha$. To form the reconstruction $\tilde{\mathbf{T}}=\Phi_m\bm\alpha$, we identify the coefficients $\bm{\alpha}$ that minimize the observation error. The general solution is given by
\begin{equation}
\label{eq:deim_a_soln}
\left(S_r^\top\Phi_m\right)^+ \mathbf{y} + \mathbf{z}=\underset{\bm{\alpha} \in \mathbb{R}^m}{\arg\min} \ \mathcal{E}_{obs}(\bm{\alpha}),  \quad \forall \mathbf{z} \in \mathcal{N}\left[S_r^\top\Phi_m\right] .
\end{equation}
The superscript $+$ denotes the Moore--Penrose pseudo-inverse and $\mathcal{N}$ represents the null space, so that $\mathcal N[S_r^\top\Phi_m]=\left\{\mathbf{z}\in\mathbb{R}^{m}: S_r^\top\Phi_m \mathbf{z} = \mathbf{0} \right\}$. We refer to $\mathbf{z}$ as the \textit{kernel vector}, an arbitrary vector within the null space of $S_r^\top\Phi_m$.  The minimizer \eqref{eq:deim_a_soln} yields the \emph{S-DEIM reconstruction}~\cite{Farazmand2024},
\begin{equation}\label{eq:SDEIM_recon}
\tilde{\mathbf{T}}(\mathbf{z}) := \Phi_m \left(S_r^\top\Phi_m\right)^+\vc y + \Phi_m \mathbf{z}, \quad \mathbf{z} \in \mathcal{N}\left[S_r^\top\Phi_m\right].
\end{equation}

DEIM and Q-DEIM use the solution with the trivial kernel vector, $\mathbf{z} = \vc 0$,  leading to the reconstruction $\tilde{\mathbf{T}}=\Phi_m \left(S_r^\top\Phi_m\right)^+\vc y$~\cite{Sorensen2010, Drmac2016}. 
In fact, when the number of sensors is greater than or equal to the number of modes ($r \geq m$), the null space is typically trivial, $\mathcal{N}\left[S_r^\top\Phi_m\right] = \{\mathbf{0}\}$, and the (Q-)DEIM reconstruction is the only choice. 
However, in the underdetermined regime, where the number of sensors is smaller than the number of modes ($r<m$), nonzero kernel vectors become an option. This regime occurs in practice when sensors are limited in number by cost and deployment obstacles, yet a larger model complexity ($m>r$) is needed to satisfactorily approximate the SST within the basis $\Phi_m$. Here, we are mainly interested in this underdetermined case. 

In this regime, one needs to answer the following question: is there an optimal kernel vector $\vc z\in\mathcal N[S_r^\top \Phi_m]$ that returns the best approximation $\tilde{\vc T}(\vc z)$ of $\vc T$?
It is straightforward to verify that the observation error~\eqref{eq:deim_err_term} is independent of the kernel vector. In other words, the instantaneous observations $\vc y(t)$ contain no additional information that would inform the choice of an optimal kernel vector~\cite{Farazmand2024}.
However,  we may ask whether there exists a kernel vector $\mathbf{z}$ which minimizes the \textit{total error},
\begin{equation}
\mathcal{E}_{tot}(\mathbf{z}(t)) := \left\|\tilde{\mathbf{T}}(\mathbf{z}(t)) - \mathbf{T}(t)\right\|.
\end{equation}
To answer this question, consider an orthonormal basis $\{\mathbf{z}_1,\ldots,\mathbf{z}_{m-r}\}$ of the null space $\mathcal{N}\left[S_r^\top\Phi_m\right]$, and define the \textit{kernel matrix} $Z := [\mathbf{z}_1 | \ldots | \mathbf{z}_{m-r}] \in \mathbb{R}^{m \times (m-r)}$. Then, there exists a unique \emph{optimal kernel vector}, 
\begin{equation}
\label{eq:opt_k}
\hat{\mathbf{z}}(t):= ZZ^\top \Phi_m^\top \mathbf{T}(t) = \underset{\mathbf{z} \in \mathcal{N}\left[S_r^\top\Phi_m\right]}{\arg\min} \mathcal{E}_{tot}(\mathbf{z}),
\end{equation}
which minimizes the total error~\cite{Farazmand2024}.
Although the kernel matrix $Z$ can be easily computed, the optimal kernel vector $\hat{\mathbf{z}}$ requires knowledge of the full SST $\mathbf{T}(t)$, and therefore is not directly computable. It appears that we might have hit an impasse again. However, as we discuss in Section~\ref{sec:recurrent}, a machine learning method can be used to estimate the optimal kernel vector from sparse observational time series. This machine learning method leverages the past history of the observations $\vc y(s)$, for $s\leq t$, in order to estimate the optimal kernel vector $\hat{\vc z}(t)$ at the current time $t$. But before discussing the machine learning method, we first review the processes for determining the basis matrix $\Phi_m$ (Section~\ref{subsec:basis}) and the selection matrix $S_r$ (Section~\ref{subsec:cpqr}).

\subsection{Basis Matrix}
\label{subsec:basis}
The basis matrix $\Phi_m$ is obtained using Proper Orthogonal Decomposition (POD) \cite{Lumey1975, Sirovich1987}, also known as Principal Component Analysis (PCA) \cite{Hotelling1933, Pearson1901}. First, we form the data matrix,
\begin{equation}
    A := [\mathbf{T}(t_1)| \mathbf{T}(t_2) | \ldots | \mathbf{T}(t_M)]\in\mathbb R^{N\times M},
\end{equation}
from the SST anomalies in the training dataset. As detailed in Section~\ref{sec:data}, each time instance $t_i$ corresponds to a week from Dec. 1989 to Dec. 2021, resulting in a total of $M=1670$ snapshots. We compute the Singular Value Decomposition (SVD) of $A$, where the $N\times N$ matrix $\Phi = [\bm\phi_1 | \bm\phi_2 |\ldots |\bm \phi_N]$ denotes the left singular matrix and $\{\bm\phi_1, \ldots, \bm\phi_N\}$ are the left singular vectors or \textit{POD modes}. We truncate the left singular matrix to the first $m$ modes in order to obtain our basis matrix $\Phi_m=\left[\bm{\phi}_{1} \vert \bm{\phi}_{2} \vert \ldots \vert \bm{\phi}_{m} \right]$. The $m$-dimensional linear subspace which best approximates the training dataset is spanned by the columns of $\Phi_m$ \cite{Koch2007}.

\begin{figure*}[htb]
\includegraphics[width=\textwidth]{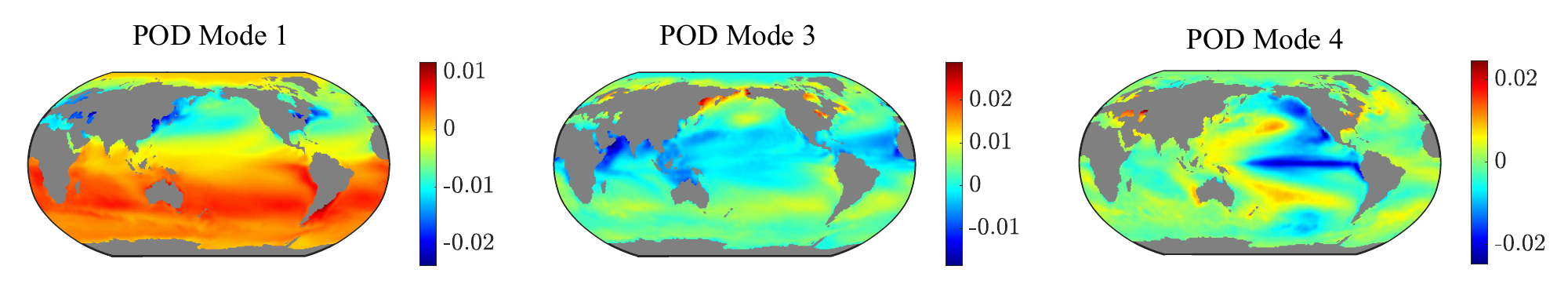}
\centering
\caption{Three POD modes generated using the training dataset.}
\label{fig:pod_modes}
\end{figure*}

Figure~\ref{fig:pod_modes} shows three POD modes, emphasizing that they often encode interpretable information about the SST. The first POD mode delineates the southern and northern hemispheres,  whereas coastline features are captured in the third mode, and the Pacific South Equatorial Current is clearly represented in the fourth mode. We emphasize that, being normalized singular vectors, the numerical values of POD modes do not represent true temperature values and should not be interpreted to indicate hot/cold temperatures; they only encode the underlying patterns within the data.

\subsection{Sensor placement}
\label{subsec:cpqr}
In our numerical results (Section~\ref{sec:results}), we consider two types of sensor placement: random and CPQR. 
The former mimics the practical situations where one does not have control over the locations of the in situ observations. The latter is performed for comparison and to demonstrate the robustness of S-DEIM regardless of the sensor placement method.

When sensor placement is random, we form the selection matrix $S_r = \left[\mathbf{e}_{i_1} \vert \mathbf{e}_{i_2} \vert \ldots \vert \mathbf{e}_{i_r} \right]$ by drawing $r$ distinct indices $\{i_1,i_2, \ldots,i_r\}$ randomly from the set of possible indices $\{1,2,\ldots,N\}$.

We also briefly review the CPQR method for sensor placement. This method was first proposed by \citeA{Drmac2016} under the assumption that the number of modes and sensors are equal ($r=m$) and was later extended to the case when $r < m$~\cite{Kakasenko2025}. 
We refer to \citeA{Drmac2016} and \citeA{Kakasenko2025} for a detailed discussion of the CPQR algorithm.

Consider the column-pivoted QR decomposition of $\Phi_r^\top$,
\begin{equation}\label{eq:cpqr}
    \Phi_r^\top\Pi = QR, \quad \Pi = \begin{bmatrix}\Pi_1 & \Pi_2\end{bmatrix}, \quad R = \begin{bmatrix}
        R_{11} & R_{12} \\ 0 & R_{22}
    \end{bmatrix},
\end{equation}
where $\Pi$ is a permutation matrix with $\Pi_1 \in \mathbb{R}^{N \times r}$ and $\Pi_2\in\mathbb{R}^{N \times (N-r)}$, $Q\in \mathbb{R}^{r \times r}$ is an orthogonal matrix, and $R_{11} \in \mathbb{R}^{r \times r}$ is an upper triangular matrix. In CPQR sensor placement, we set the selection matrix $S_r = \Pi_1$, the first $r$ columns of the permutation matrix $\Pi$. In equation~\eqref{eq:cpqr}, we used the decomposition of the matrix $\Phi_r$, the first $r$ columns of the full basis matrix $\Phi_m$. This choice is purely based on the empirical observation that it returned more accurate reconstructions compared to CPQR sensor placement based on $\Phi_m$.

If the selection matrix $S_r$ is obtained from the CPQR decomposition, the resulting matrix $S_r^\top\Phi_m$ has full rank~\cite{gu1996}, so that $\text{dim}(\mathcal{N}[S_r^\top\Phi_m])= m-r$. However, when the sensor locations are randomly selected, $S_r^\top\Phi_m$ is not guaranteed to have full rank. Nonetheless, we observed that $S_r^\top\Phi_m$ tends to be full-rank in practice, even when the sensors are placed randomly.

Finally, we highlight an implicit assumption in our sensor placement methods, namely that the sensor locations are static. This allows in situ observations from moored buoys but excludes observations from drifters and ships that move over time. Moving sensors would imply a time-dependent selection matrix $S_r(t)$, which in turn implies a time-dependent kernel matrix $Z(t)$. The current S-DEIM theory~\cite{Farazmand2024, Farazmand2024b} is not suitable for such time-dependent cases whose handling requires further theoretical developments.

\section{S-DEIM with Recurrent Neural Networks}\label{sec:recurrent}
We recall that the S-DEIM reconstructions $\tilde{\vc T}(\vc z)$ are most accurate when the optimal kernel vector $\vc z = \vc{\hat z}$ is used; see Eq.~\eqref{eq:opt_k}. But computing the optimal kernel vector $\hat{ \mathbf{z}}$ requires knowledge of the entire temperature field $\mathbf{T}$, which is unavailable.  Therefore, the feasibility of S-DEIM for SST reconstruction hinges on our ability to estimate the optimal kernel vector from sparse in situ data $\vc y$.
To this end, we utilize recurrent neural networks (RNNs) to learn the mapping from the in situ observational time series to the optimal kernel vector. An RNN is a type of machine learning model particularly suited for sequential time series data. Its internal memory allows it to learn the time-dependent patterns linking the time series of sparse sensor readings to the optimal kernel coordinates. The RNN takes a time series of past observations,
\begin{equation}
    \mathcal{Y}{(t)} = \{ \mathbf{y}(s) : s \leq t \},
\end{equation}
as inputs to predict the optimal kernel vector $\hat{\mathbf{z}}(t)$ as its output. As discussed in \citeA{Farazmand2024b}, it is necessary to use the past history of in situ observations, since instantaneous observations $\vc y(t)$ alone do not contain enough information to infer the optimal kernel vector $\hat{\vc z}(t)$.

As depicted in Figure~\ref{fig:my_pipeline}, our framework consists of an offline training stage and a real-time reconstruction stage. The offline stage uses historical high-fidelity SST data to prepare the necessary components for S-DEIM reconstruction and for training the RNN. In particular, in the offline stage, we compute the POD basis matrix ($\Phi_m$) and the selection matrix ($S_r$) for sensor placement. Furthermore, we use the high-fidelity SST data to prepare the training pairs, which consists of input time series $\mathcal{Y}_{tr}(t)$ and the corresponding target output $\hat{\vc{z}}(t)$. We use these input-output pairs to train the RNN. Once trained, the RNN can be deployed for real-time reconstruction. In this stage, the pretrained RNN takes the available sparse in situ observations $\mathcal{Y}(t)$ to predict the kernel vector ${\vc{z}}(t)$. This predicted vector is then used for the S-DEIM reconstruction~\eqref{eq:SDEIM_recon} of the SST.

An important practical note is in order here. The RNN can, in principle, be trained to predict the $m$-dimensional optimal kernel vector $\hat{\mathbf{z}}(t)\in\mathcal{N}[S_r^\top\Phi_m]\subset \mathbb R^m$. However, without loss of generality, the dimension of the RNN output can be decreased, thus improving the performance of the RNN~\cite{Farazmand2024b}.
To this end, recall that the columns of the kernel matrix ${Z}$ form an orthonormal basis for the null space $\mathcal{N}[S_r^\top\Phi_m]$. Therefore, we can write $\hat{\vc z}(t) = Z\hat{\bm\xi}(t)$, where the unique vector
$\hat{\bm{\xi}}(t) \in \mathbb{R}^{m-r}$,
\begin{equation}
\label{eq:xi}
\hat{\bm{\xi}}(t) = Z^\top\hat{\mathbf{z}}(t) ={Z}^\top{\Phi}_m^\top\mathbf{T}(t),
\end{equation}
denotes the coordinates of $\hat{\mathbf{z}}(t) \in \mathbb{R}^m$ in the basis $Z$. 
Since the dimension of the coordinate vector $\hat{\bm{\xi}}(t) \in \mathbb{R}^{m-r}$ is smaller than that of the full kernel vector $\hat{\mathbf{z}}(t) \in \mathbb{R}^{m}$, it is more efficient to train the network to predict $\hat{\bm{\xi}}(t)$. Subsequently, the kernel vector $\hat{\vc z}(t) = Z\hat{\bm\xi}(t)$ can be recovered uniquely from the kernel coefficients $\hat{\bm\xi}(t)$.

In Section~\ref{sec:results}, we report the SST reconstruction results for two types of RNNs: the Reservoir Computing (RC) network and the Long Short-Term Memory (LSTM) network. The architecture of each network and the hyperparameter tuning process are detailed in~\ref{app:rnn_architectures}. The eventual hyperparameters that we used for the RNNs are detailed in Table~\ref{tab:rnn_params}.
\begin{table*}[]
\centering
\begin{tabular}{|l|c||l|c|} 
\hline
\multicolumn{2}{|c||}{\textbf{RC}} & \multicolumn{2}{c|}{\textbf{LSTM}} \\
\hline
\textbf{Parameter} & \textbf{Value} & \textbf{Parameter} & \textbf{Value} \\
\hline
Reservoir Size $(N_r)$ & 100 & Hidden Units $(h_t)$ & 300 \\
Learning Rate ($\alpha$) & 1.0 & Initial Learning Rate ($\alpha$) & 0.01 \\
Network Sparsity ($s$) & 40\% & Drop Period ($\tau)$& 50 \\
Regularization Parameter $(\lambda$) & $10^{-8}$ & Drop Rate ($\gamma$) & 0.1 \\
\hline
\end{tabular}
\caption{Hyperparameters for the two types of RNN architectures used in this paper: RC and LSTM.}
\label{tab:rnn_params}
\end{table*}

Table~\ref{tab:rnn_time} reports the computational time it takes to train each RNN and the inference time on the test dataset, i.e., computational time it takes for the pretrained RNN to produce SST reconstructions from in situ observations. The computations are carried out on a workstation with a Windows 11 operating system, Intel Core i9-14900K CPUs, and 32 GB memory.
\begin{table}[h!]
\centering
\begin{tabular}{|l|c|c|} 
\hline
& \textbf{Training} & \textbf{Inference}\\
\hline
\textbf{RC} & 0.01 & 0.03  \\ \hline
\textbf{LSTM} & 61.15 & 0.32  \\
\hline
\end{tabular}
\caption{Wall clock time for training and inference for each RNN architecture. The units of time are in seconds.}
\label{tab:rnn_time}
\end{table}

For the LSTM network, the training process takes approximately one minute, whereas for the RC network the training time is significantly less, around 0.01 seconds. The inference time for both network architectures is less than a second.

\section{Results and Discussion}
\label{sec:results}
In this section, we demonstrate the effectiveness of S-DEIM in reconstructing SST snapshots from partial observations. We form S-DEIM reconstructions using two types of RNNs (RC and LSTM) to compute the kernel vector. We compare these results against Q-DEIM reconstructions, which use the trivial kernel vector $\vc z=\vc 0$. In addition, we report the S-DEIM reconstruction using the optimal kernel vector $\hat{\vc z}$ as well as the best fit approximation $\hat{\vc T}$. Recall that the S-DEIM reconstruction with the optimal kernel vector and the best fit approximation are not computable in practice, since they require access to the full SST. Nevertheless, they are reported here since they provide a comparison baseline for the best possible approximations.

We use the weekly NOAA Optimal Interpolation SST dataset (OISST) spanning the period from December 1989 to January 2023~\cite{Reynolds2002}. As outlined in Section~\ref{sec:data}, we divide the dataset into two subsets: training (December 1989 - December 2021) and testing (January 2022 - January 2023). Each SST snapshot is given on a $360\times180$ grid, recording the SST in increments of one degree. We vectorize each snapshot, discarding the grid points that fall on land. In addition, we center the data by removing the temporal mean of the training data from all SST snapshots. The reported relative errors refer to the relative error of a reconstruction $\tilde{\mathbf{T}}$ compared to the truth $\mathbf{T}$, 
\begin{equation}
RE = \frac{\|\tilde{\mathbf{T}} - \mathbf{T}\|}{\|\mathbf{T}\|},
\label{eq:rel_error}
\end{equation}
where all temperature vectors refer to mean-removed SST (i.e., SST anomalies).

Throughout this section, the reconstructions $\tilde{\vc T}$ are obtained from $r=100$ in situ observations, which represents a very sparse distribution; roughly 0.2\% of the high-resolution grid. The number of sensors $r=100$ is close to the number of NOAA's moored buoys available in the National Data Buoy Center (NDBC)~\cite{McCall1998}.

As the quality of the reconstruction is dependent on the sensor locations, we also compare the reconstructions for different methods of sensor selection. Section~\ref{subsec:opt_sp} discusses the SST reconstructions from CPQR sensor placement. In Section~\ref{subsec:rand_sp}, we also report results from random sensor placement, demonstrating the robustness of S-DEIM reconstructions. In particular, we demonstrate that S-DEIM is not as sensitive to sensor placement method as DEIM. Note that the SST data used for our computations is derived from observational data which inherently include noise. Even so, adding up to 10\% additional synthetic Gaussian noise to the observations did not change the results appreciably. In Section~\ref{Comparisons}, we compare the S-DEIM results to reconstructions from a convolutional neural network (CNN) with Voronoi tessellation~\cite{Fukami2021}. Notably, LSTM-based S-DEIM generates slightly more accurate reconstructions with a significantly shorter training time than the CNN.

\subsection{Reconstruction from CPQR Sensor Placement}
\label{subsec:opt_sp}
In this section, we present the S-DEIM reconstructions resulting from CPQR sensor placement (see Section~\ref{subsec:cpqr}). We first examine the reconstruction error for a fixed number of sensors, $r=100$, as the number of modes $m$ vary, and the reconstruction error for a fixed number of modes $m=300$, as the number of sensors vary. Then, we analyze the temporal and spatial distribution of the errors.

\begin{figure*}
    \centering
    \includegraphics[width=1\linewidth]{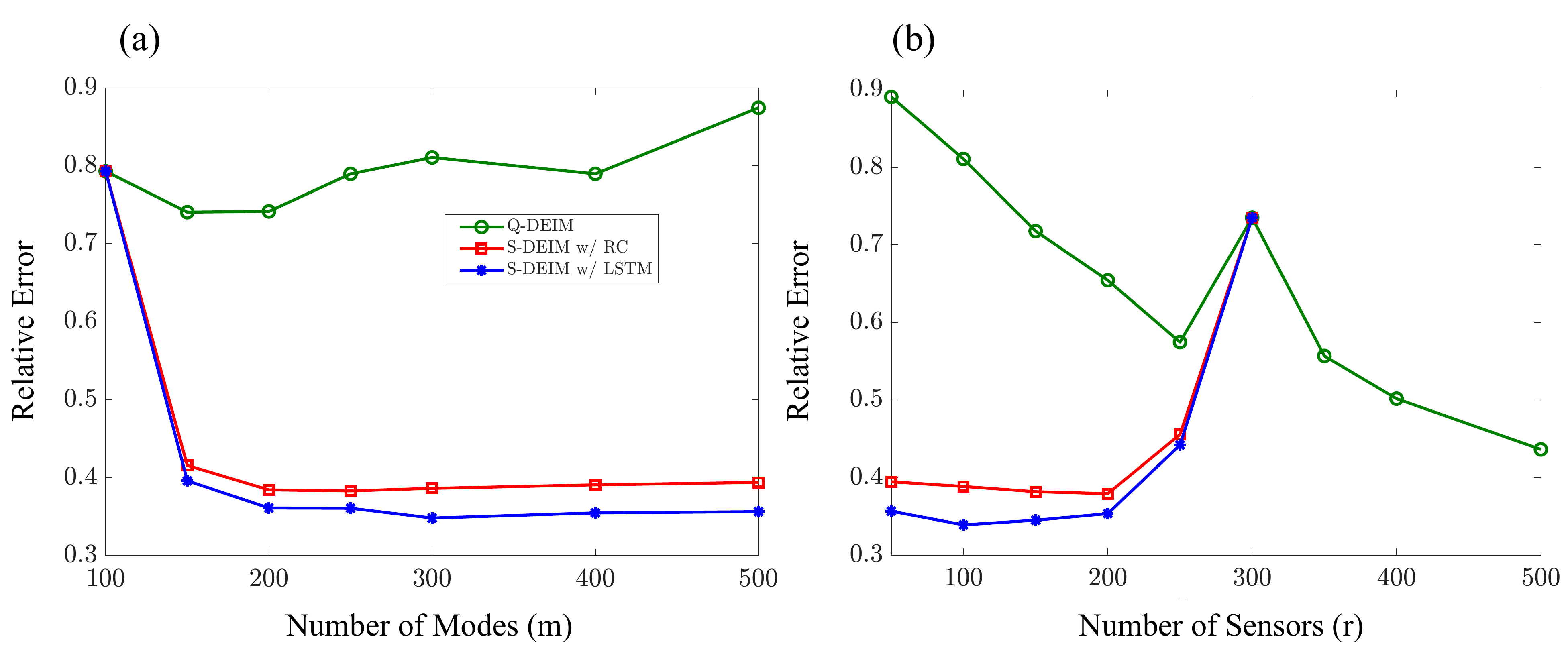}
    \caption{Mean of relative reconstruction error for (a) an increasing number of modes $m$ with fixed number of sensors $r=100$ and (b) an increasing number of sensors $r$ with fixed number of modes $m=300$. The mean is taken over the snapshots in the test dataset.}
    \label{fig:mean_std_modes}
\end{figure*}

\emph{Varying the number of modes and sensors}: Figure~\ref{fig:mean_std_modes}(a) shows the mean relative error of the reconstructions, computed from the test data (Jan. 2022--Jan. 2023), as the number of modes increases with $r=100$ sensors. The relative error of the Q-DEIM reconstructions, which use the trivial kernel vector $\vc z=\vc 0$, fluctuates near $80\%$. As the number of modes increases, the error decreases slightly at first, but then increases for $m>200$, so that no particular trend is observed for the Q-DEIM reconstruction error.

Next we turn to the S-DEIM reconstructions using RC and LSTM neural networks. Note that when $m=100$, the number of modes matches the number of sensors, resulting in a trivial null space $\mathcal N[S_r^\top\Phi_m]=\{\vc 0\}$. Therefore, the Q-DEIM and S-DEIM reconstructions are equivalent when $m=r=100$. However, when $m>r$ the S-DEIM reconstructions significantly outperform Q-DEIM.
More specifically, we observe an immediate drop in the error of the S-DEIM reconstructions, from $79.30\%$  with $m=100$ modes to approximately $40\%$ with $m=150$ modes. The error then continues to decline to approximately $36\%$ as the number of modes increases from 150 to 300.
At this point, the S-DEIM errors begin to plateau; further increasing the number of modes, beyond $m=300$, does not result in significantly higher accuracy.
\begin{figure*}
    \centering
    \includegraphics[width=\textwidth]{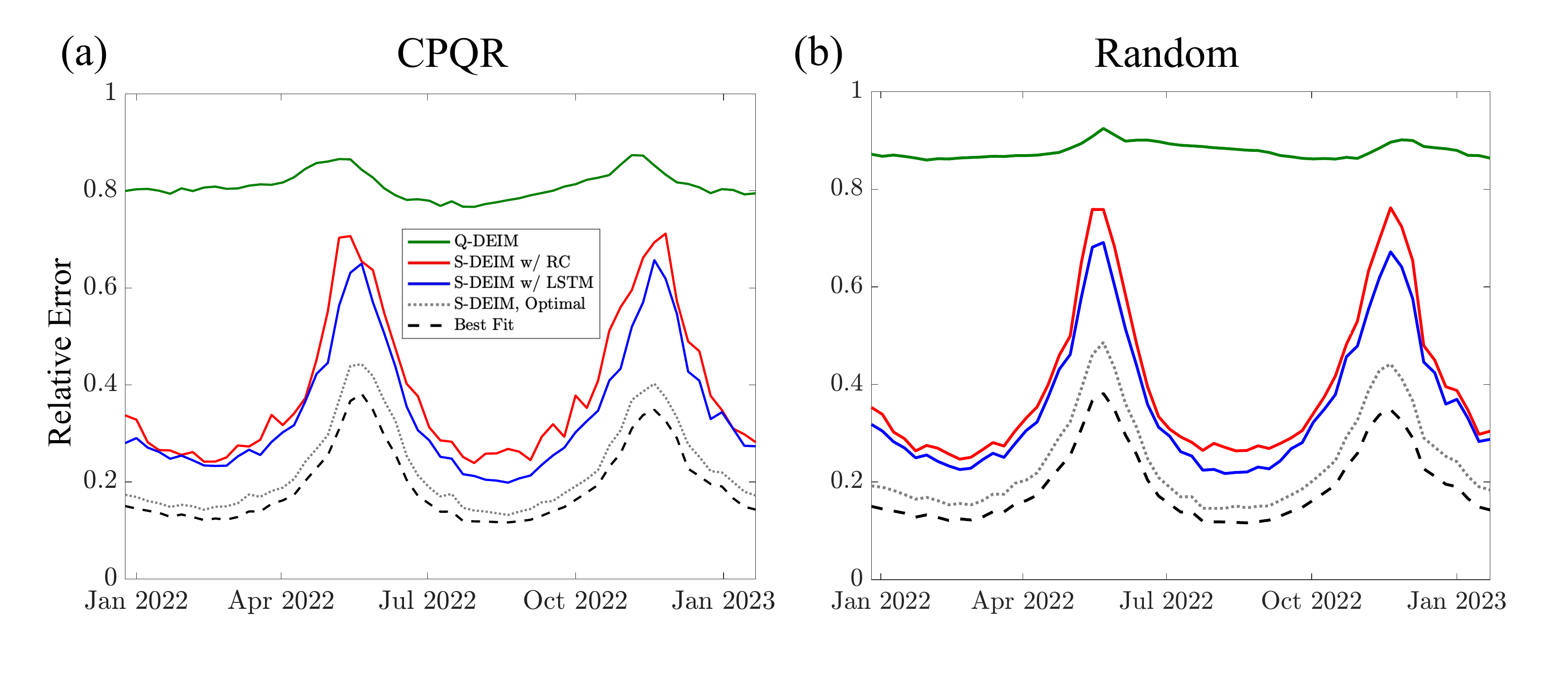}
    \caption{Relative reconstruction error with $r=100$ sensors and $m=300$ modes. (a) CPQR sensor placement. (b) Randomly placed sensors.}
    \label{fig:sensors}
\end{figure*}

Fixing the number of modes at $m=300$, Figure~\ref{fig:mean_std_modes}(b) shows the mean relative error of the reconstructions as the number of sensors $r$ increases. Initially the Q-DEIM reconstruction error decreases as the number of sensors increases, since the additional observations provide more information for estimating the SST field. When the number of sensors is equal to the number of modes, the Q-DEIM reconstruction error spikes; it is known that, for $m=r$, Q-DEIM is more sensitive to observational noise~\cite{Brunton2021,Drmac2020}. Increasing the number of sensors beyond the number of modes leads to oversampling, which improves reconstruction accuracy again.

In contrast, the S-DEIM reconstruction error remains relatively constant from $r=50$ to $r=200$ sensors and well below the corresponding Q-DEIM reconstruction errors. As the number of sensors increases, the dimension of the kernel vector decreases, leaving S-DEIM with progressively fewer degrees of freedom in which to improve upon the Q-DEIM reconstruction. At the same time, as the dimension of the null space decreases, it is also easier for the RNN to learn the optimal kernel vector; as such, we do not observe a significant change in the S-DEIM accuracy for $r\leq 200$. As the number of sensors approaches $r=300$, the S-DEIM reconstruction errors sharply increase, converging to the Q-DEIM error. Recall that for $r\geq m=300$ sensors, the null space is trivial, so that the S-DEIM and Q-DEIM reconstructions coincide.

Due to diminishing returns in performance and the increasing computational cost of going beyond 300 modes, we fix the number of modes at $m=300$ to further analyze the S-DEIM reconstructions. Although Q-DEIM is usually performed with an equal number of sensors and modes, here the Q-DEIM error for $m=100$ is similar to the error for $m=300$ (see Figure \ref{fig:mean_std_modes}(a)).
Likewise, we fix the number of sensors at $r=100$ as that is when we observe the best performance of LSTM-based S-DEIM. 

\emph{Temporal evolution:} Figure \ref{fig:sensors}(a) shows the weekly reconstruction error for the test data (Jan. 2022-Jan. 2023). The S-DEIM reconstructions are significantly better than Q-DEIM for every time instance. Specifically, the LSTM- and RC-based S-DEIM reconstructions achieve mean relative errors of $34.82\%$ and  $38.64\%$, respectively, which are more than $40\%$ lower than Q-DEIM’s error of  $81.08\%$. We attribute the superior accuracy of the LSTM-based reconstructions to the increased complexity of the LSTM network. The LSTM network has 541,400 total trainable parameters compared to only 20,000 for the RC network. The best fit and optimal S-DEIM reconstruction errors are also shown in Figure~\ref{fig:sensors} for reference, but recall that they are not computable from sparse observational data. All reconstruction errors peak during seasonal transitions in May-June and November-December. In particular,  the RNN-based S-DEIM reconstruction errors peak at approximately $65\%$ during these months, which is significantly higher than their minima of approximately 25\% during the months of February-March and August-September. This suggests that the seasonal shifts may require a higher number of POD modes to accurately capture. 

It is noteworthy that there is an appreciable gap between the S-DEIM reconstructions using RNNs and the optimal S-DEIM reconstructions. In theory, this gap can be closed with more complex network architectures. But in practice, this gap is expected due to training error (inability to find the global loss minimizer using numerical optimization) and the limited expressivity of a finite network~\cite{Bottou2007,welper2025}

\begin{figure}[h]
    \centering
    \includegraphics[width=0.5\textwidth]{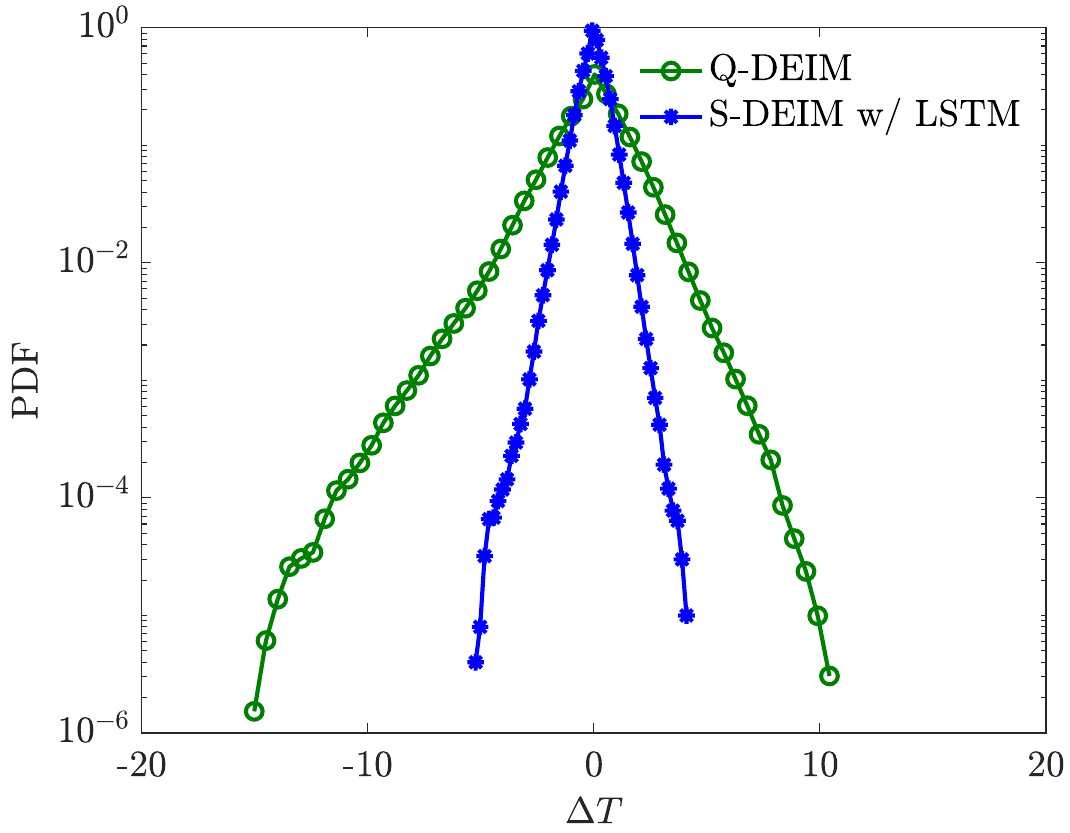}
    \caption{The PDF of the difference $\Delta T$ between the reconstructed SST and the truth for Q-DEIM and S-DEIM. For S-DEIM, $\Delta T$ is concentrated around $0^\circ$C, with $91\%$ of reconstructions falling within $\pm 1^\circ$C. For Q-DEIM, $\Delta T$ exhibits heavier tails and only $51\%$ are within $\pm 1^\circ$C.}
    \label{fig:diff_pdf}
\end{figure}

\emph{Spatial distribution:} Next, we consider the variability in reconstruction errors between S-DEIM and Q-DEIM. We compute the difference $\Delta T$ between the reconstruction $\tilde T$ and the true SST $T$ at each grid point $\mathbf{x}_i$ and each time instance $t_j$, where $\Delta T(\mathbf{x}_i,t_j)=\tilde{T}(\mathbf{x}_i,t_j)-T(\mathbf{x}_i,t_j).$ In Figure~\ref{fig:diff_pdf}, we compare the Probability Density Function (PDF) of $\Delta T$ for Q-DEIM and LSTM-based S-DEIM. The PDF for the RC-based S-DEIM is nearly identical to that of the LSTM-based S-DEIM and is therefore omitted. The S-DEIM error $\Delta T$ exhibits significantly smaller variance, with a narrower range from approximately $-5^\circ$C to $4^\circ$C. In contrast, the Q-DEIM errors span a wider range, from around $-15^\circ$C to $10^\circ$C. Furthermore, the S-DEIM errors are highly concentrated near zero, with $91\%$ falling within $\pm 1^\circ$C of the truth. In contrast, only $51\%$ of Q-DEIM reconstructions are within this range. Hence, S-DEIM is much less prone to extreme errors in both frequency and intensity compared to Q-DEIM.

\begin{figure*}[h]
	\centering
	\includegraphics[width=.85\textwidth]{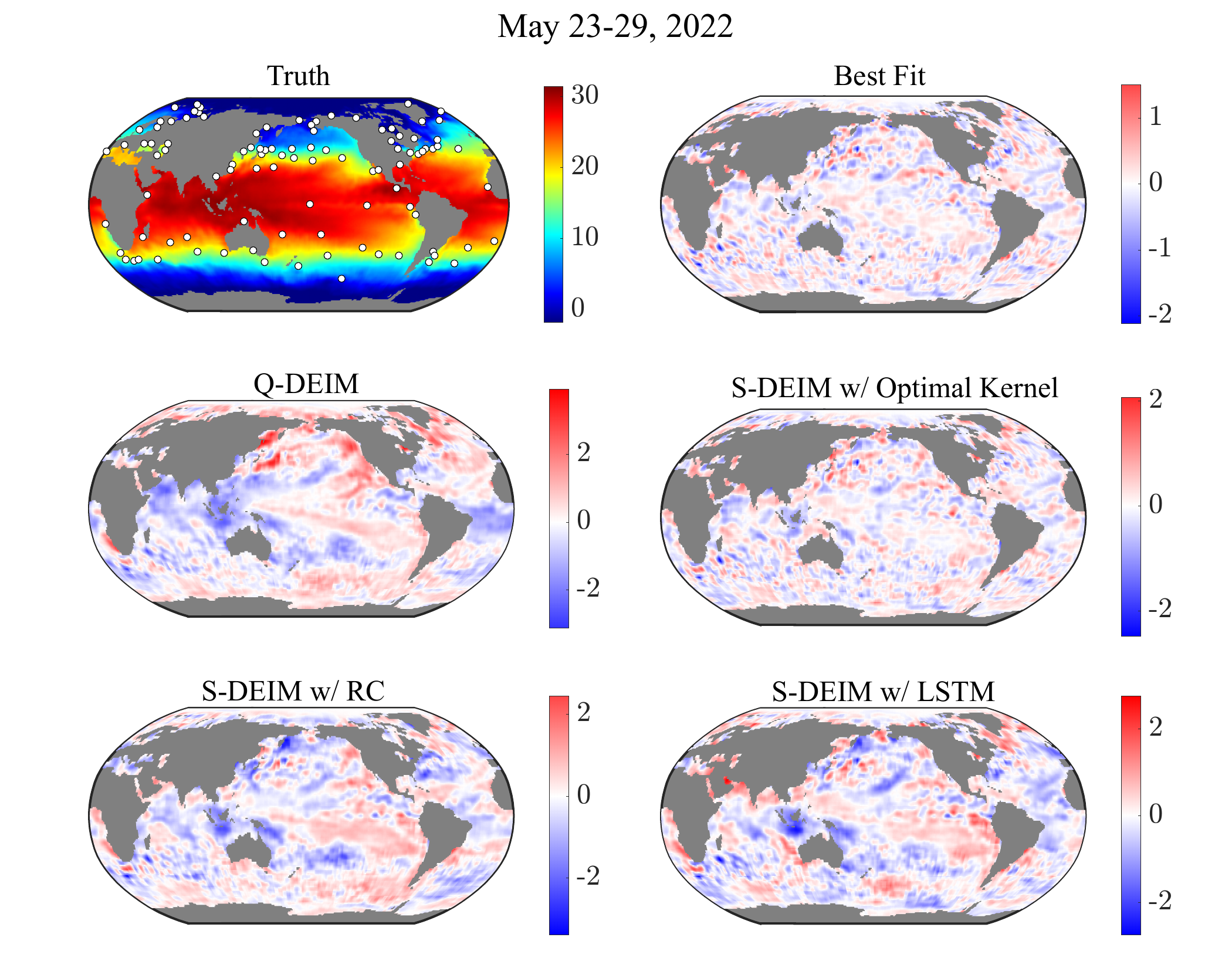}
	\caption{True SST for the week of May 23-29, 2022, together with sensor locations marked by white circles (top left). Other panels show the difference $\Delta T$ between each reconstruction and the true SST. May 23-29, 2022 is the week in which the S-DEIM reconstructions have maximum relative error.
	}
	\label{fig:recon_week21}
\end{figure*}

\begin{figure*}[h]
	\centering
	\includegraphics[width=.85\textwidth]{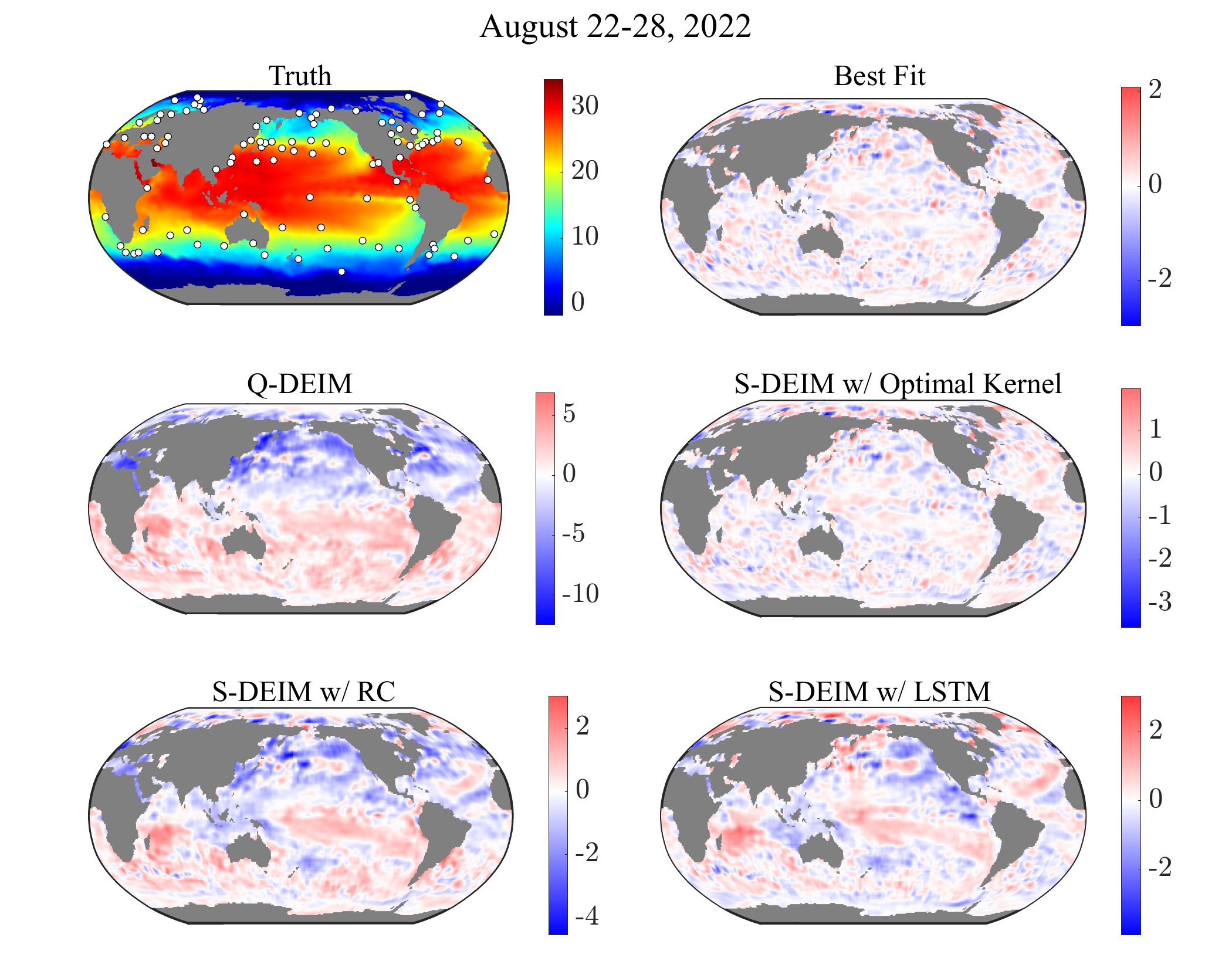}
	\caption{True SST for the week of August 22-28, 2022 together with sensor locations marked by white circles (top left). Other panels show the difference $\Delta T$ between each reconstruction and the true SST. August 22-28, 2022 is the week in which the S-DEIM reconstructions have minimum relative error.
	}
	\label{fig:recon_week34}
\end{figure*}

We select two representative weeks to showcase the spatial distribution of error in our reconstructions. Figures~\ref{fig:recon_week21} and~\ref{fig:recon_week34} compare the reconstructions to the true SST during the weeks of maximum and minimum relative error, respectively. More precisely, Figure~\ref{fig:recon_week21} corresponds to the week of May 23-29, 2022, and Figure~\ref{fig:recon_week34} corresponds to August 22-28, 2022.
The upper left panel in each figure displays the true SST field (not anomalies) as well as the CPQR sensor locations marked with white circles. The remaining panels show the difference $\Delta T$ between the reconstructed field $\tilde{T}$ and the true temperature field $T$, i.e., $\Delta T=\tilde{T} - T$. In \ref{app:sst_anom}, we also show the SST anomalies for completeness.

The top right panel is the best fit approximation, and the center right panel shows the optimal S-DEIM reconstruction. These two are only included for reference, since they are not computable from sparse observations. 
The optimal S-DEIM and best fit approximations show small and spatially dispersed errors, whereas the RC (bottom left) and LSTM (bottom right) S-DEIM  reconstructions exhibit a more structured error distribution.
For example, in Figure~\ref{fig:recon_week21}, both RNN-based S-DEIM reconstructions underestimate the SST in the Indonesian region by approximately $2^\circ$C. Figure~\ref{fig:recon_week34} shows a similar underestimation across the Western Pacific Ocean.  More generally, across all snapshots, we found a few locations where the maximum S-DEIM reconstruction errors tend to cluster: along the Kuroshio current, along the Agulhas Current, and along the Pacific Ocean's South Equatorial Current. In contrast, the Q-DEIM reconstruction errors are more uniformly distributed around the globe (see, e.g., center-left panel of Figure~\ref{fig:recon_week34}). 

In summary, with CPQR sensor placement, S-DEIM returns significantly better reconstructions compared to Q-DEIM. The computational cost of S-DEIM is only slightly higher than Q-DEIM. The primary computational cost lies in the one-time offline training of the RNN. Once trained, the S-DEIM inference of the SST takes less than a second; see Table~\ref{tab:rnn_time}.

\subsection{Reconstruction from Random Sensor Placement}
\label{subsec:rand_sp}
In practice, the location of in situ observations is determined based on logistical considerations as opposed to the optimality of the resulting SST reconstructions. To emulate this practical situation, we consider SST reconstructions from randomly placed sensors. More precisely, we consider $25$ realizations of $r=100$ sensors, placed at random points on the sea surface. For each realization, we compute DEIM and S-DEIM reconstructions and assess the sensitivity of different reconstruction methods to the sensor placement method. 

As discussed in Section~\ref{subsec:cpqr}, when the sensors are placed randomly, the matrix $S_r^{\top}\Phi_m$ may be rank deficient. In such cases, the dimension of the null space $\mathcal{N}[S_r^{\top}\Phi_m]$ is greater than $m-r$. To ensure that the dimension of the null space stays consistent, we discard selection matrices $S_r$ for which $S_r^{\top}\Phi_m$ does not have full rank. This guarantees that $\text{rank}(S_r^\top\Phi_m)=r$ across all 25 realizations of random sensors. 
\begin{figure}
    \centering
    \includegraphics[width=0.5\textwidth]{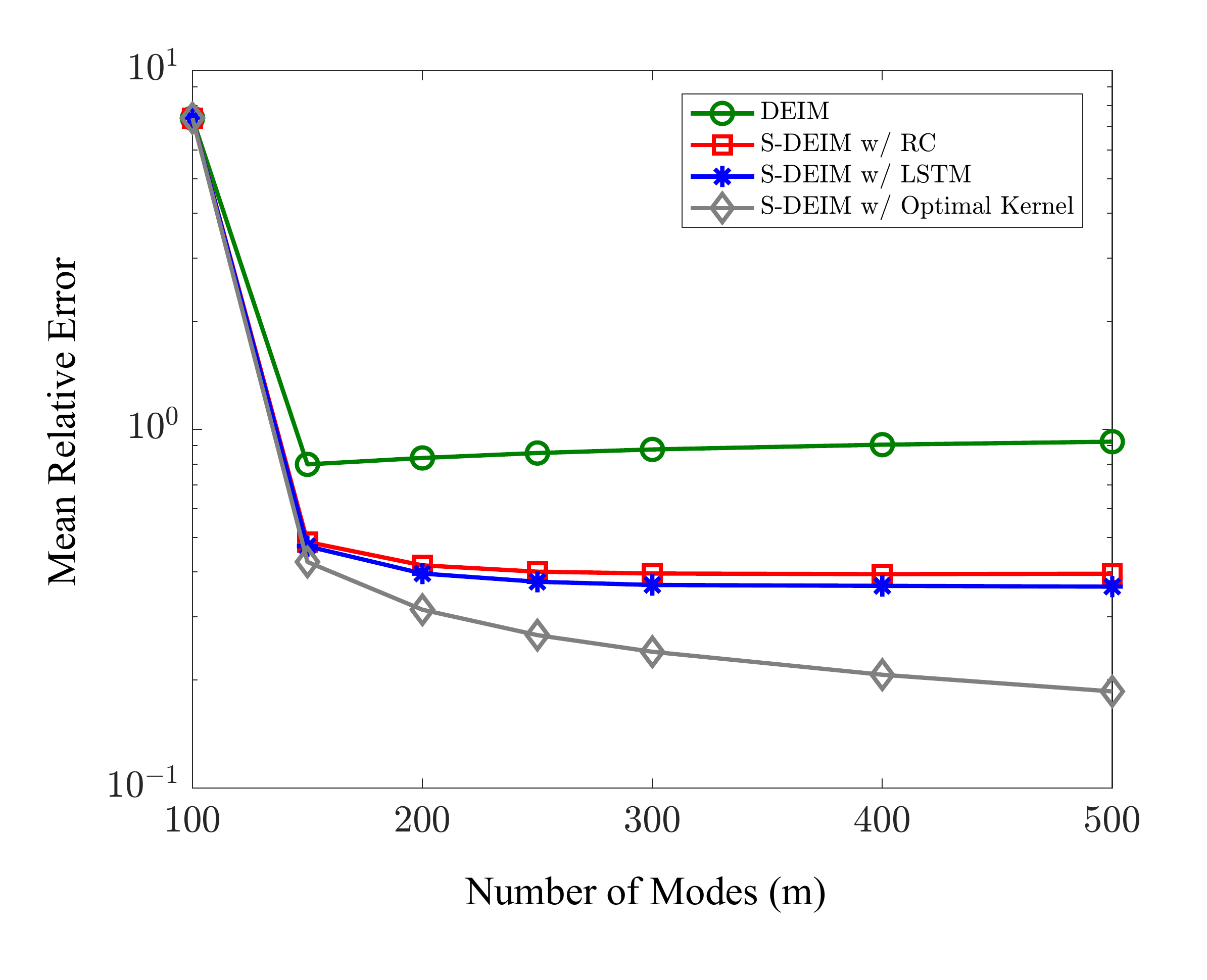}
    \caption{Mean reconstruction error averaged over 25 different realizations of randomly placed sensors as a function of the number of modes.}
    \label{fig:modes_randsensors}
\end{figure}

Figure~\ref{fig:modes_randsensors} shows the mean relative error, where the average is taken over all test data (Jan. 2022--Jan. 2023) and all 25 realizations of randomly placed sensors. We plot the mean relative error as a function of the number of modes $m$. This figure is analogous to Fig.~\ref{fig:mean_std_modes} where the sensors were placed based on the CPQR algorithm. We first discuss the DEIM reconstructions. 
For the same number of sensors and modes $(m=r=100)$, DEIM performs very poorly with an average relative error of approximately $737\%$.  Compare this to the Q-DEIM error of 79.30\% in Fig.~\ref{fig:mean_std_modes}, where CPQR sensor placement was used. This observation is in line with previous studies that report similarly large errors for DEIM when sensors are randomly placed~\cite{Manohar2018}. 

Interestingly, as the number of modes increases from $m=100$ to $m=150$, there is a sharp drop in average relative error for all reconstruction methods. In particular, the DEIM error drops an order of magnitude to $79.83\%$, which is similar to the Q-DEIM reconstruction errors with CPQR sensor placement. In summary, DEIM reconstructions from randomly placed sensors are rather abysmal when the same number of sensors and modes is used. However, when the number of modes is increased while keeping the number of sensors fixed, the DEIM reconstructions with random sensor placement are almost as accurate as the Q-DEIM reconstructions with CPQR sensor placement. 

Now we turn our attention to the S-DEIM reconstructions. Recall that for $m=r$, S-DEIM and DEIM reconstructions coincide; both use the trivial kernel vector $\vc z=\vc 0$. However, when $m>r$, the two differ in that S-DEIM uses an estimation of the optimal kernel vector $\hat{\vc z}$.
In Figure~\ref{fig:modes_randsensors}, we observe that the S-DEIM reconstruction error decays rapidly as soon as the number of modes is larger than the number of sensors. We also note that S-DEIM reconstructions are significantly more accurate than DEIM reconstructions for any $m>r=100$; S-DEIM errors are roughly $40\%$ lower than DEIM errors. The S-DEIM error eventually plateaus when $m\geq 300$, where further increasing the number of modes does not lead to significant gains in terms of accuracy. Interestingly, the S-DEIM reconstruction errors are comparable to the case where CPQR sensor placement was used (cf. Figure~\ref{fig:mean_std_modes}). This indicates that S-DEIM reconstructions are not sensitive to the sensor placement method. 
\begin{table}[h!]
	\centering
\begin{tabular}{|l|c|c|}
\hline
\multicolumn{1}{|c|}{\textbf{Method}} & \textbf{Mean RE}  & \textbf{Maximum RE} \\ \hline
Q-DEIM (DEIM)                                            & 0.8108 (0.8787)                       & 0.8736 (0.9246)                          \\ 
S-DEIM w/ RC                                        & 0.3864 (0.3962)                       & 0.7128 (0.7621)                          \\
S-DEIM w/ LSTM                                      & 0.3482 (0.3680)                       & 0.6367 (0.6927)                         \\
S-DEIM w/ Opt. Kernel                            & 0.2256 (0.2397)                       & 0.4427 (0.4862)                          \\
\hline
\end{tabular}
\caption{Mean relative error and maximum relative error for various reconstruction methods with $m=300$ modes and $r=100$ sensors. The errors corresponding to CPQR sensor placement are reported next to the errors corresponding to random sensor placement (in parenthesis).}
\label{error_table}
\end{table}

To see this more clearly, in Table~\ref{error_table}, we report the mean and maximum relative error of the reconstructions for $m=300$ modes.
The mean is obtained by averaging the error over the entire test dataset. Similarly, the maximum is taken over the entire duration of the test dataset (Jan. 2022-Jan. 2023). DEIM with random sensor placement performs about $7\%$ worse on average than Q-DEIM, which uses CPQR sensor placement. For randomly placed sensors, the mean relative error for S-DEIM with optimal kernel vector, RC, and LSTM are $23.97\%, 39.62\%$, and $36.80\%$, respectively. These errors are only about 1 to 2\% higher than the mean relative error of S-DEIM with CPQR sensor placement. 

Figure~\ref{fig:sensors}(b) shows the temporal evolution of the relative error over the period of Jan. 2022--Jan. 2023 (test dataset). 
The general trends are very similar to the reconstructions with CPQR sensor placement. In particular, maximum errors occur during the months of May and November, where seasonal changes make predictions more challenging. These maximum errors are also reported in Table~\ref{error_table}.
Changing the sensor placement method, from CPQR to random, leads to an increase in the maximum error. However, these increases are modest at about 4 to 5\% for both DEIM and S-DEIM.

\subsection{Comparison with a Voronoi-CNN}\label{Comparisons}
    We further compare the performance of S-DEIM against reconstructions obtained by a convolutional neural network. 
 	In particular, we use the Voronoi tessellation-assisted convolutional neural network (Voronoi-CNN), proposed by \citeA{Fukami2021}. Given a set of sensor locations, a Voronoi tessellation of the spatial domain is constructed so that each point in the domain is assigned the SST anomaly of its nearest sensor according to the Euclidean distance. All the points that map to a specific sensor form a Voronoi cell. The observed value at each sensor is then extended to all points in the corresponding Voronoi cell, producing a piecewise constant approximation of the full SST field from sparse observations. This Voronoi field is the input for the CNN, which is trained to recover an estimate for the corresponding high-resolution SST field.

     We use the open-source implementation provided by~\citeA{Fukami2021}, including their Voronoi prepossessing procedure and CNN architecture, as described in~\ref{app:CNN_Arch}. The Voronoi-CNN is trained and evaluated using the same $r=100$ CPQR sensor locations used in Section~\ref{subsec:opt_sp}.

        \begin{figure}
            \centering
            \includegraphics[width=0.5\linewidth]{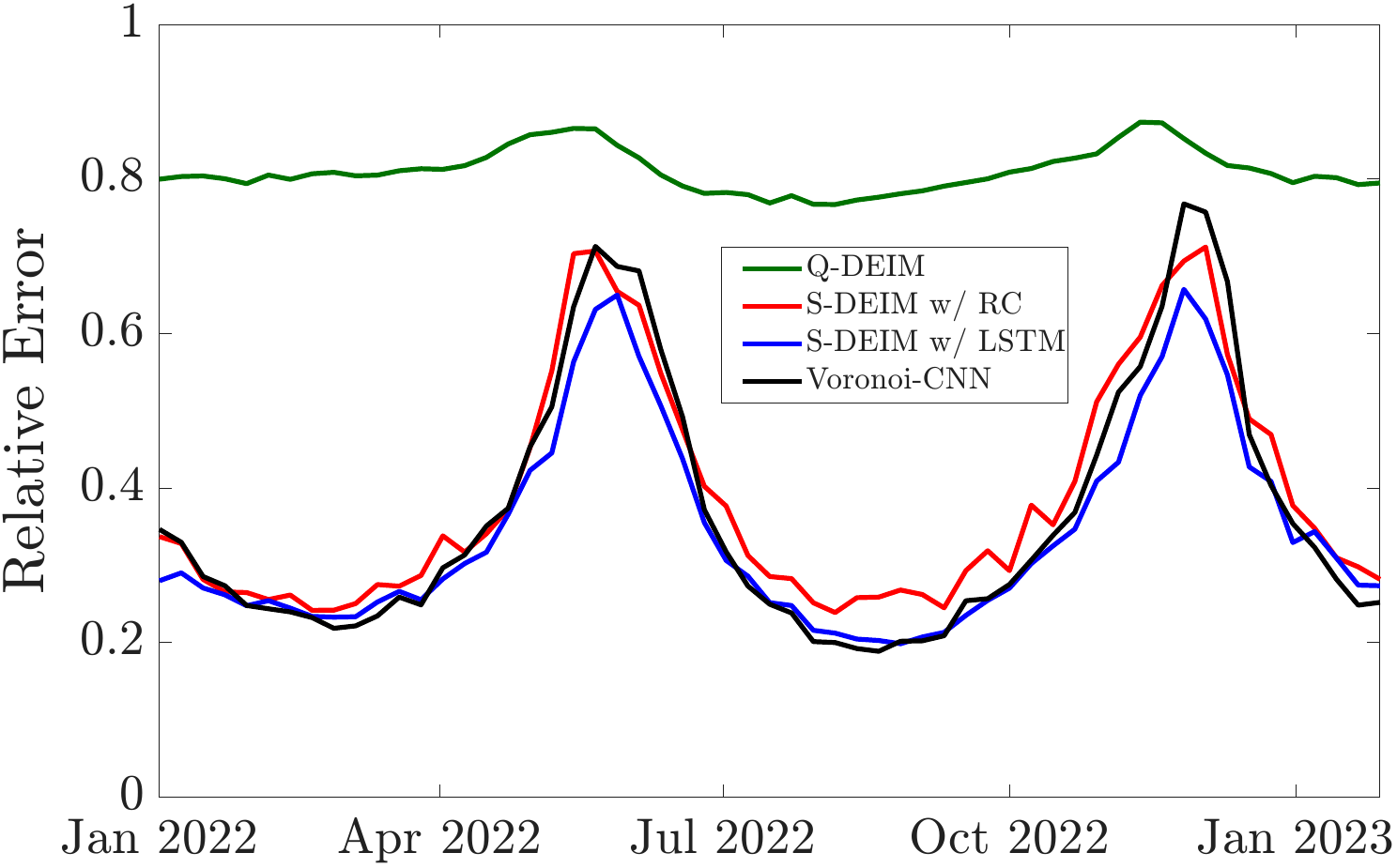}
            \caption{Relative reconstruction error for Q-DEIM, S-DEIM with RC, S-DEIM with LSTM, and the Voronoi-CNN baseline. All methods use the same $r=100$ CPQR sensor locations. The Q-DEIM and S-DEIM reconstructions use $m=300$ POD modes.}
            \label{fig:CnnErrors}
        \end{figure}
        
       The resulting reconstruction errors are shown in Figure~\ref{fig:CnnErrors}, indicating an overall behavior similar to S-DEIM reconstructions. More specifically, the Voronoi-CNN achieves an average relative reconstruction error of $36.48\%$, compared with $38.65\%$ for S-DEIM with RC and $34.34\%$ for S-DEIM with LSTM. Although the Voronoi-CNN substantially improves over Q-DEIM ($81.08\%$ error), it does not outperform the LSTM-based S-DEIM reconstructions. In addition, the offline computational cost of the Voronoi-CNN is substantially higher than that of the S-DEIM. Training the CNN required $5904$ seconds ($1.64$ hours), whereas the training time for the RNN-based S-DEIM is at most $61.15$ seconds; see Table~\ref{tab:rnn_time}. The inference times were comparable: the Voronoi-CNN requires approximately $0.39$ seconds, while the S-DEIM inference times are at most $0.32$ seconds. Therefore, S-DEIM remains competitive with a CNN-based sparse reconstruction method, in spite of its simplicity, interpretability, and significantly lower training time. We also point out that the Voronoi tessellation seems crucial to the success of CNN, since the CNN results without tessellation were significantly less accurate, with relative errors near 100\% (not shown here for brevity).

\section{Conclusions}\label{sec:conclusion}
Data-driven estimation of geophysical variables, such as sea surface temperature, remains a challenging task, especially when only a sparse set of observations is available. Here, we showed that Sparse Discrete Empirical Interpolation Method (S-DEIM), with the aid of recurrent neural networks, can significantly improve SST reconstructions from sparse observations. More specifically, the relative error of S-DEIM is 40\% lower than its predecessors: the DEIM and Q-DEIM methods. Furthermore, 91\% of S-DEIM estimates fall within $\pm 1^\circ$C of the true SST. This number is only 51\% for Q-DEIM. Furthermore, we compared the S-DEIM results against the Voronoi tessellation-assisted CNN of \citeA{Fukami2021}. Although Voronoi-CNN reconstructions are comparable to S-DEIM in terms of accuracy, training the CNN takes two orders of magnitude longer compared to S-DEIM. In addition, S-DEIM is simpler and more interpretable compared to the Voronoi-CNN network.

We demonstrated that the S-DEIM reconstructions are not unduly sensitive to sensor placement, i.e., the locations from which in situ observations are gathered. For example, S-DEIM reconstruction errors increased by only 1-2\% when switching from a nearly optimal sensor placement method (i.e., CPQR) to randomly placed sensors. S-DEIM is also robust to moderate amounts of observational noise. For example, adding up to 5\% Gaussian noise to the in situ observations did not change the reconstructions appreciably. 

The success of S-DEIM hinges on the use of a kernel vector, which is often ignored in empirical interpolation. The optimal value of this kernel vector is not computable from instantaneous in situ observations. Here, we used RNNs to estimate the optimal kernel vector from time series of past observations.
Although this returned satisfactory results, there exists an approximately 10\% gap between the accuracy of optimal S-DEIM and RNN-based S-DEIM reconstructions. 
Closing this gap could be the subject of future work. In theory, one might expect that this gap can be closed with more training data and/or increasing network complexity. But in practice, increasing the network complexity (number of layers and nodes) of the RNNs did not close this gap; neither did increasing the number of epochs in the numerical optimization (see Appendix~\ref{app:LSTM_arch}). Furthermore, varying the length of the training data between 800 weeks and 1670 weeks (not shown here) did not appreciably affect the reconstruction errors, indicating that increasing the length of the training data beyond 1670 weeks will likely not close the gap with optimal S-DEIM either. Therefore, future work should consider utilizing alternative RNN architectures, such as transformers \cite{fu2024, song2024}, to explore the possibility of improving upon our results. Another scenario that should be explored is the possible existence of a fundamental obstruction which leads to the observed 10\% gap in performance between RNN-based S-DEIM and optimal S-DEIM.

Another direction for future work consists of reconstruction from multi-fidelity sensors~\cite{Brunton2021}. S-DEIM assumes that each observation is of equal quality.  However, measurements collected from different sources have varying levels of uncertainty. One may address this problem, for example, by introducing a weighted norm in the least squares problem~\eqref{eq:deim_a_soln} underlying S-DEIM, such that lower quality observations contribute less to the residual~\eqref{eq:deim_err_term}.

Finally, in S-DEIM it is assumed that the sensors are static, i.e., they do not move over time. This limits us to using in situ observations from moored buoys. To enable the use of observations from drifters and ships, further theoretical developments are needed to generalize S-DEIM for use with moving sensors. 

\section*{Conflict of Interest disclosure}
The authors have no conflicts to disclose.

\section*{Data availability statement}
The sea surface temperature data is publicly available from the National Oceanic and Atmospheric Administration (NOAA) website: \url{https://www.ncei.noaa.gov/products/optimum-interpolation-sst}. This data and the accompanying codes are publicly available at \citeA{All2026_zenodo}.

\acknowledgments
This work was partially supported by the National Science Foundation (NSF) through the grant DMS-2220548 (Algorithms for Threat Detection Program). C.A., K.H., M.M., C.N., and M.F. were supported by NSF's REU program, through the award DMS-2349611. L.B.E. was supported by NSF's RTG program, through the award DMS-2342344.

\appendix
\section{Recurrent Neural Network Architectures}
\label{app:rnn_architectures}

We provide a high-level overview of the two Recurrent Neural Network (RNN) architectures used in this study, namely the Reservoir Computing (RC) network and the Long Short-Term Memory (LSTM) network. We also discuss the hyperparameter tuning that was undertaken in this study.

We pre-process the data by vectorizing the SST anomaly snapshots and removing the points that fall on land. The RNN input at each time step is the sparse observation vector $\vc y(t)\in \mathbb{R}^r$,
\begin{equation}
\vc y(t)=S_r^\top \vc{T}(t).
\end{equation}
The training input is the full historical sequence of sparse sensor observations over the training period. For $r=100$ sensors, the input matrix corresponds to a \(100 \times 1670\) matrix for 100 sensor observations over 1670 weekly training snapshots. The target at each time instance is the optimal kernel coordinates $\hat{\pmb\xi}(t)\in\mathbb{R}^{m-r}$,
\begin{equation}
\hat{\pmb\xi}(t)=Z^\top \Phi_m^\top \vc T(t).
\end{equation}
The networks are trained using the full input and output sequences from the training dataset.
\subsection{Reservoir Computing (RC)}
\begin{figure}
	\centering
	\includegraphics[width=0.5\textwidth]{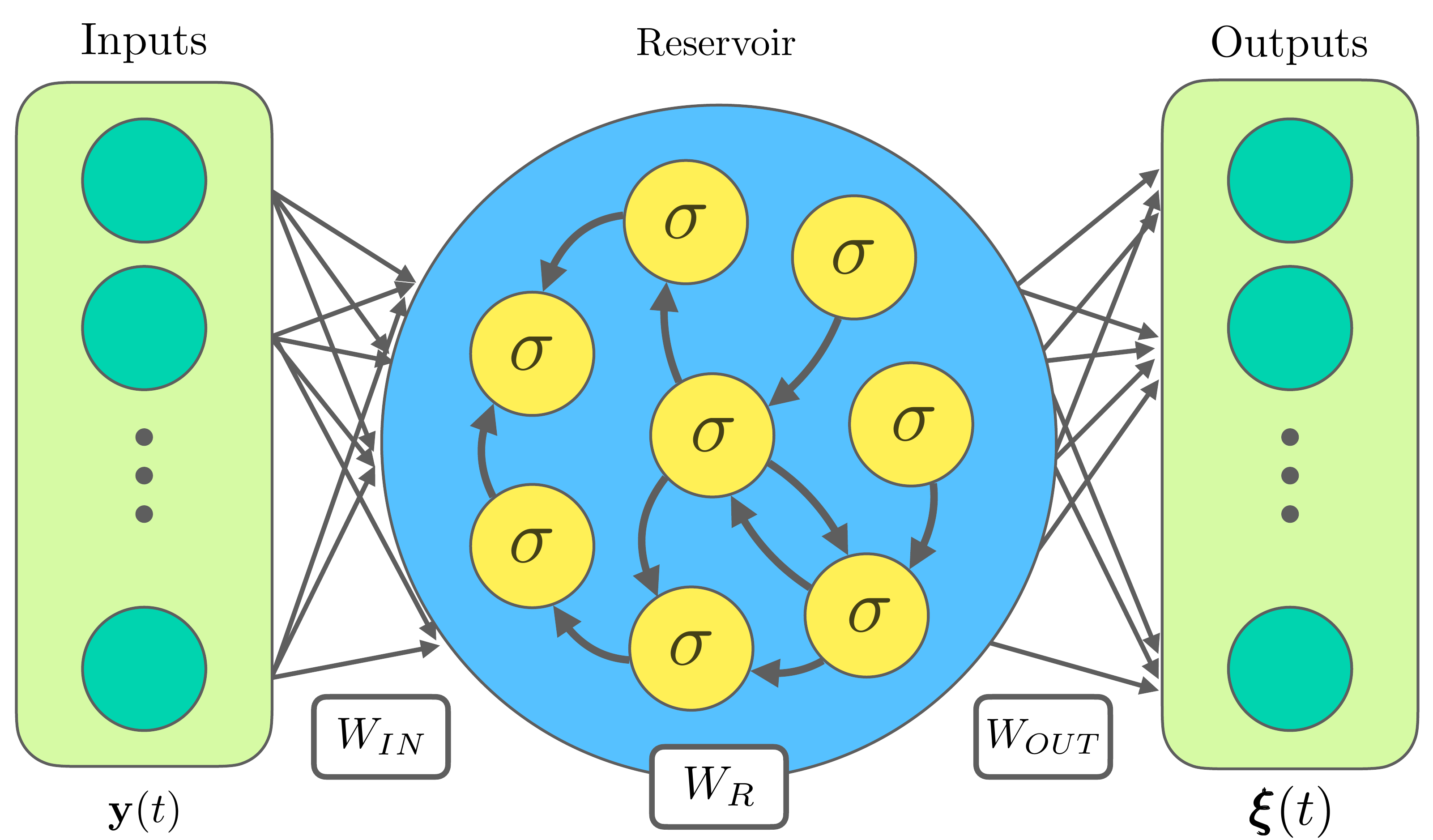}
	\caption{The RC network consists of an input layer with weights $W_{IN}$, a fixed random reservoir with internal weights $W_{R}$, and a trainable output layer with weights $W_{OUT}$.}
	\label{fig:rc_arch}
\end{figure}

Reservoir Computing is a specialized type of RNN notable for its training efficiency. The RC architecture consists of an input layer, a reservoir, and a trainable output layer, as depicted in Figure \ref{fig:rc_arch}. At any time $t_i$, we denote the reservoir state by $\mathbf{r}(t_i) \in \mathbb{R}^{N_r}$ where $N_r \gg \max\{r, m\}$ is the reservoir size, $r$ denotes the number of sensors, and $m$ denotes the number of modes. The input data, the time series of observations $\mathbf{y}(t)$, is lifted to the high-dimensional state space of the reservoir via the {input weight matrix} $W_{IN} \in \mathbb{R}^{N_r \times r}$. The reservoir dynamics is regulated by its internal weights $W_{R} \in \mathbb{R}^{N_r \times N_r}$ and biases $\mathbf{b} \in \mathbb{R}^{N_r}$. 

At each time step $t_i$, the reservoir state is updated with the equation,
\begin{equation}
    \mathbf{r}(t_{i+1}) = (1-\alpha){\mathbf{r}}(t_i) + \alpha \sigma(W_R \mathbf{r}(t_i) + W_{IN} \mathbf{y}(t_i) + \mathbf{b}),
    \label{eq:rc_update}
\end{equation}
where $\alpha$ is the learning rate, and $\sigma(x)=\tanh(x)$ is the activation function applied element-wise.

A defining characteristic of the RC methodology is that the weight matrices for the input layer ($W_{IN}$), the reservoir ($W_{R}$), and the bias vector ($\mathbf{b}$) are all initialized randomly and remain fixed. We refer to \citeA{Jaeger2009} and \citeA{Farazmand2024b} for details concerning the initialization of these random quantities. The only component of the network that is trained is the {output weight matrix} $W_{OUT} \in \mathbb{R}^{(m-r) \times N_r}$. This layer performs a linear mapping from the high-dimensional reservoir state $\mathbf{r}(t)$ to the target output, here the optimal kernel coordinates ${\hat{\bm{\xi}}}(t)$. 

We collect the reservoir states $\mathbf{r}(t_i)$ into the matrix $R$ and the optimal kernel coordinates $\hat{\boldsymbol{\xi}}(t_i)$ $\Xi$ into another matrix $\Xi$,
\begin{equation}
R = [\mathbf{r}(t_1)|\mathbf{r}(t_2)|\cdots|\mathbf{r}(t_M)], \qquad \Xi = [\hat{\boldsymbol{\xi}}(t_1)|\hat{\boldsymbol{\xi}}(t_2)|\cdots|\hat{\boldsymbol{\xi}}(t_M)].
\end{equation}
The optimal output weights $W_{OUT}$ are determined by solving the regularized least-squares problem,
\begin{equation}
    W_{OUT} = \underset{W}{\arg\min} \|WR - \Xi\|_{F}^{2} + \lambda\|W\|_{F}^{2},
\end{equation}
where $0 < \lambda \ll 1$ is a regularization parameter and $\|\cdot\|_{F}$ denotes the Frobenius norm. The corresponding solution to the regularized least-squares problem is given by
\begin{equation}
    W_{OUT} = {\Xi} {R}^T ({R}{R}^T + \lambda {I})^{-1}.
    \label{eq:wout_solution}
\end{equation}
Training an RC network is computationally inexpensive (see Table~\ref{tab:rnn_time}) because only the output layer $W_{OUT}$ needs to be learned, and its optimal value is known in closed form~\eqref{eq:wout_solution}. Notably, the RC network does not use a numerical optimization algorithm, early stopping, or normalization of the inputs and outputs.

To determine the optimal architecture for this study, we performed a hyperparameter search over several key parameters. The search space includes the learning rate $\alpha \in \{0.2, 0.4, 0.8, 1.0\}$, the reservoir size $N_r \in \{100, 250, 500, 1000\}$, and the sparsity of the reservoir weight matrix $W_R$, $s \in \{10\%, 20\%, 40\%\}$. The final hyperparameters that we use for this paper, determined by the smallest reconstruction relative error on the training data, are detailed in Table~\ref{tab:rnn_params}.

\subsection{Long Short-Term Memory (LSTM)}\label{app:LSTM_arch}
\begin{figure*}
	\centering
	\includegraphics[width=0.65\textwidth]{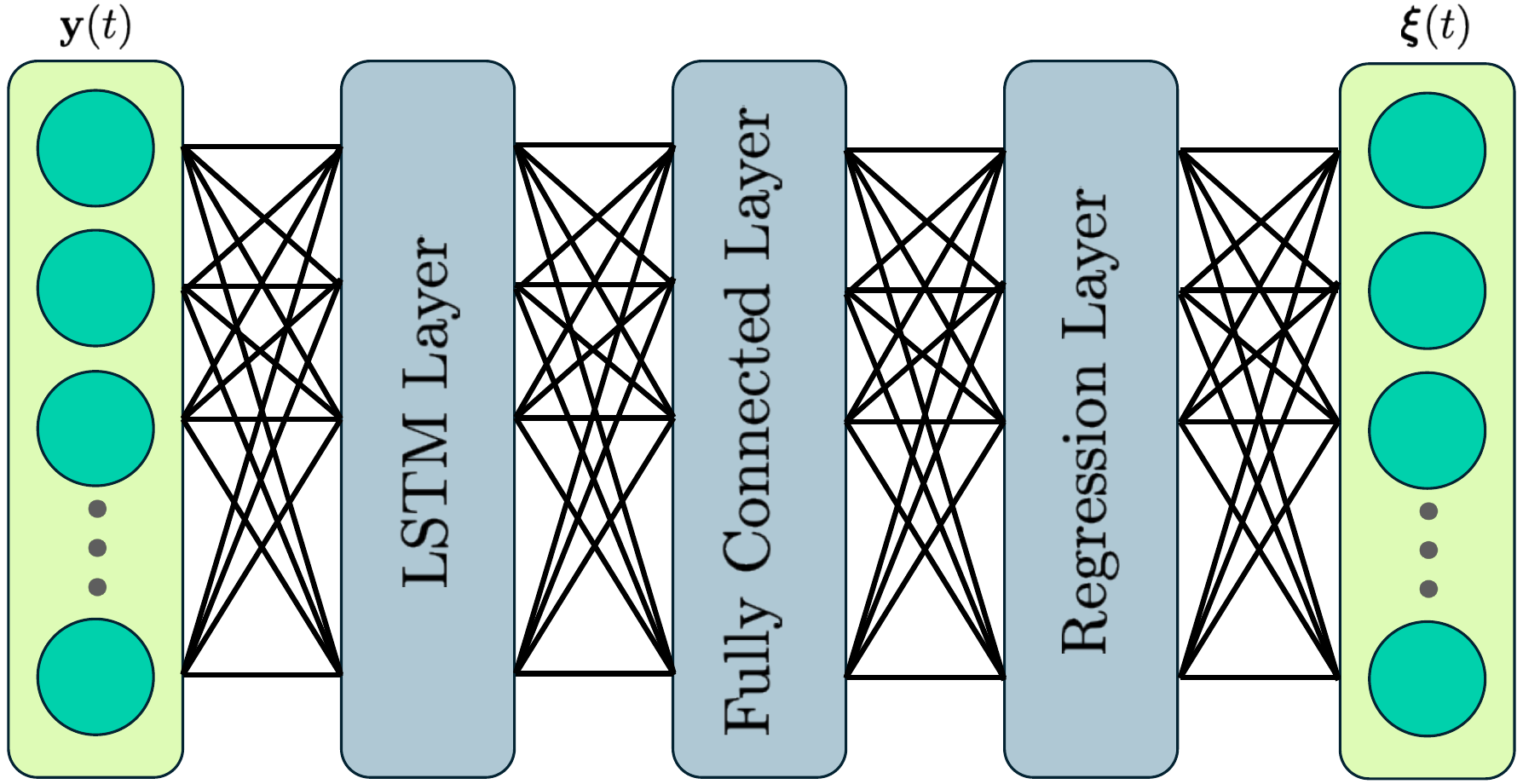}
	\caption{Architecture of the Long Short-Term Memory (LSTM) model. The LSTM network architecture processes input data $\mathbf{y}(t)$ through an LSTM network followed by a fully connected layer and a regression layer to predict the output vector $\bm{\xi}(t)$.}
	\label{fig:LSTM_arch}
\end{figure*}

Long Short-Term Memory (LSTM) networks are a more complex and powerful type of RNN designed specifically to overcome challenges with learning long-term dependencies in data~\cite{Hochreiter1997,VanHoudt2020}. Figure \ref{fig:LSTM_arch} depicts the architecture we use to learn the kernel coordinates $\bm\xi(t)$. 

The network processes the input vector of in situ observations $\mathbf{y}(t)$. This input is first fed into an LSTM layer composed of memory cells that use gating mechanisms (input, forget, and output gates) to regulate the flow of information. The hidden state from the LSTM layer is then passed to a single-layer fully connected network with $m-r=200$ nodes, which translates the nonlinear features extracted by the LSTM into continuous numerical values. The fully connected layer has a linear activation function, so that $\pmb\xi = W \vc h +\vc b$ where $W$ and $\vc b$ are the weights and biases of the fully connected layer, respectively. In addition, $\vc h$ denotes the output of the LSTM layer and $\pmb\xi$ denotes the output of the entire network.
The role of the regression layer is merely to compute the loss function, 
\begin{equation}
\mathcal{L}_{\mathrm{LSTM}}
=
\frac{1}{2S}
\sum_{n=1}^{S}
\left\|
\pmb\xi(t_n)
-
\hat{\pmb\xi}(t_n)
\right\|_2^2,
\end{equation}
where \(S=1670\) is the training sequence length, and \(t_n\) denotes the \(n\)-th weekly training snapshot. Here, $\pmb\xi$ denotes the output of the entire network and $\hat{\pmb\xi}$ denotes the target output (i.e., the truth).

The LSTM inputs and targets are normalized before training using the mean and standard deviation of the training data. The LSTM is trained using the Adam optimizer over 300 epochs. Increasing the epochs beyond 300 did not improve the results appreciably.  Early stopping was not used in training this network.

To optimize the predictive performance of the LSTM network, we conducted hyperparameter tuning. The parameters investigated included the number of hidden units $h_t \in \{100, 200, 300\}$, the initial learning rates $\alpha \in \{10^{-2}, 5 \times 10^{-3}, 10^{-3}\}$, the learning rate drop period after $\tau \in \{50, 100\}$ epochs, and the corresponding learning rate drop factors $\gamma\in \{0.1, 0.5\}$. The optimal hyperparameters were chosen based on the combination of hyperparameters that yielded the lowest reconstruction error on the training data. These parameters are reported in Table~\ref{tab:rnn_params}. 

\begin{figure}
    \centering
    \includegraphics[width=.5\textwidth]{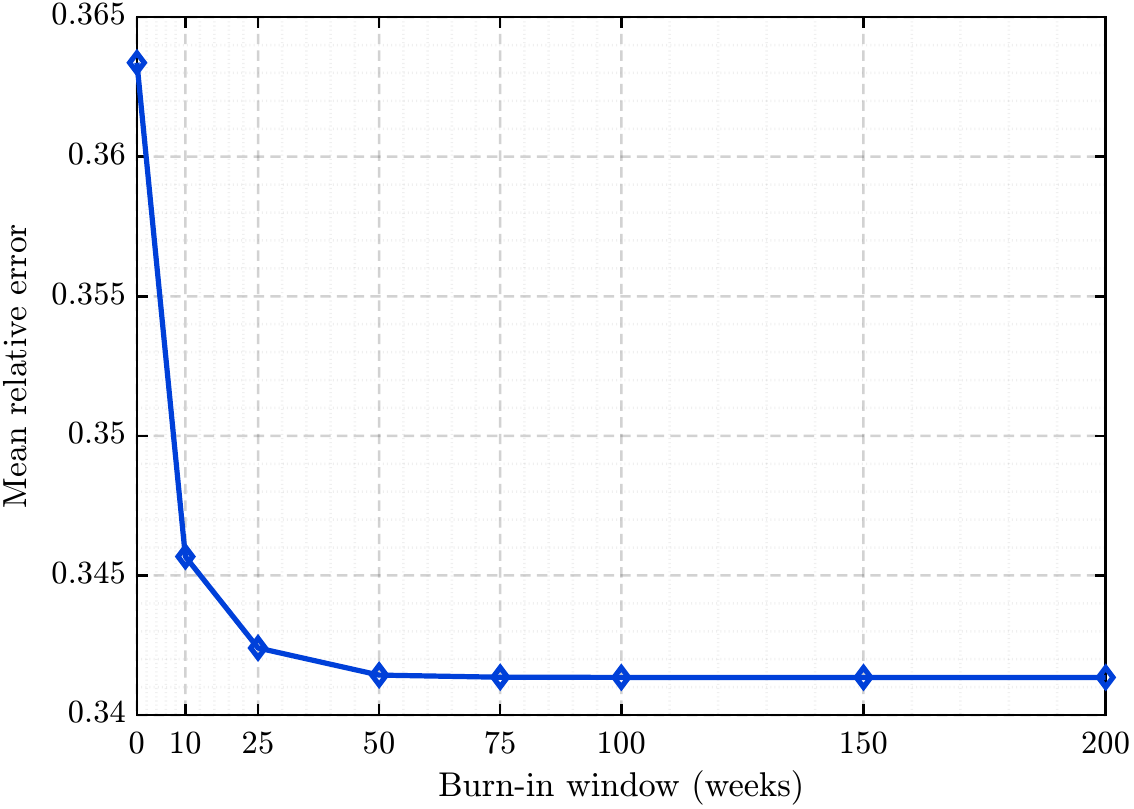}
    \caption{Mean reconstruction error of LSTM-based S-DEIM over different burn-in windows. The LSTM is trained with 300 modes and 100 sensors.}
    \label{fig:burnin_test}
    \end{figure}

During testing, a burn-in (or washout) period from the end of the training dataset is fed into the RNN as input before making predictions on the test set. The burn-in window allows the RNN to sync with the observational input data before reconstruction of test data begins~\cite{Jaeger2002,Beintema2021}. A burn-in window of 50 weeks is chosen here after performing a sweep across a range of burn-in windows, as shown in Figure~\ref{fig:burnin_test}. The relative error steadily decreases until 50 weeks, at which point additional burn-in resulted in negligible improvement in performance.
    
\section{SST Anomalies}\label{app:sst_anom}
\begin{figure*}[h]
	\centering
	\includegraphics[width=.85\textwidth]{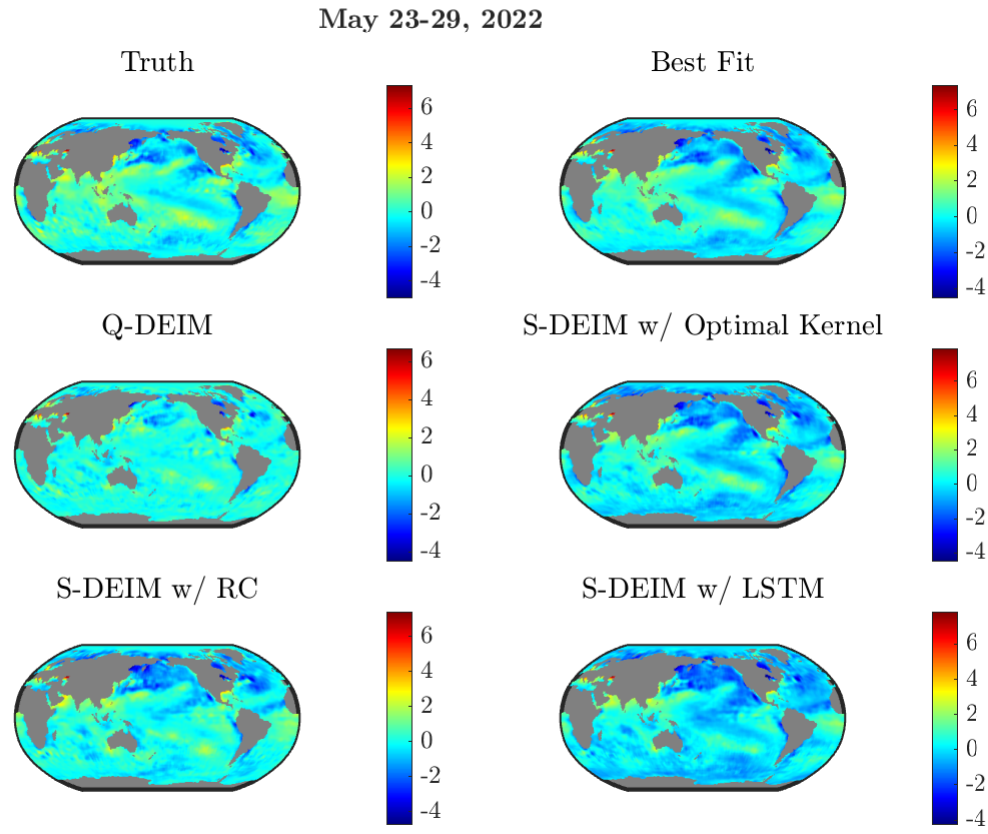}
	\caption{Same as Figure~\ref{fig:recon_week21}, except that here all panels show the SST anomalies. May 23-29, 2022 is the week in which the S-DEIM reconstructions have maximum relative error.
	}
	\label{fig:recon_week21_anomaly}
\end{figure*}

\begin{figure*}[h]
	\centering
	\includegraphics[width=.85\textwidth]{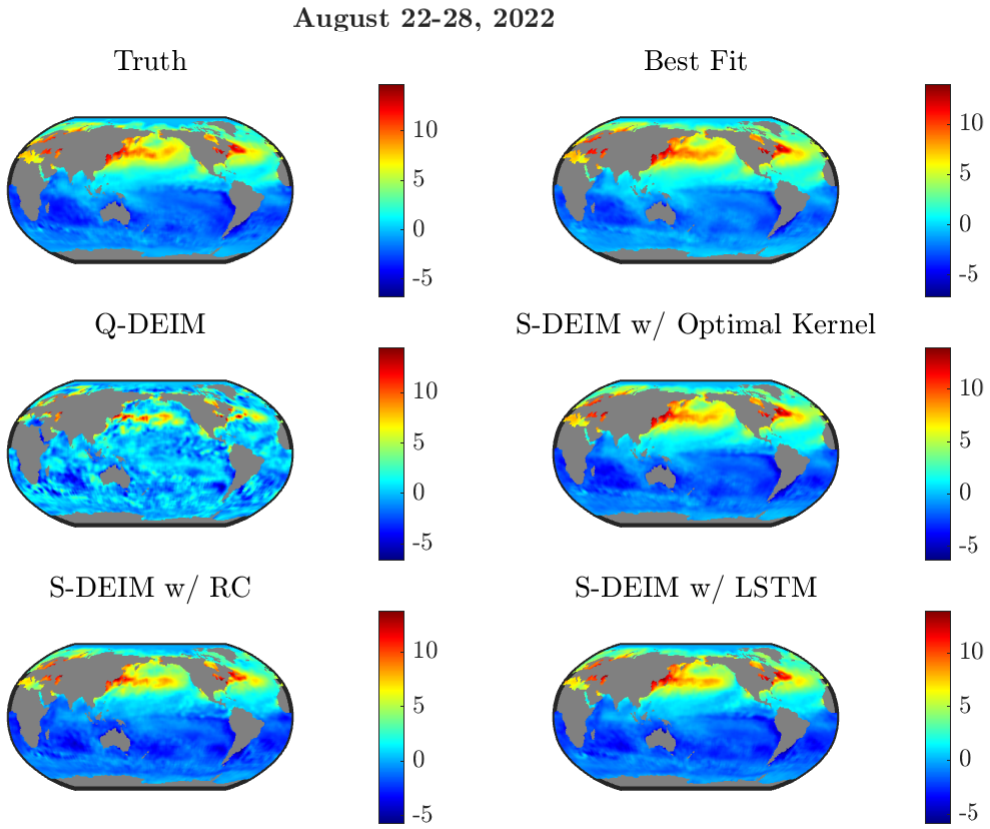}
	\caption{Same as Figure~\ref{fig:recon_week34}, except that here all panels show the SST anomalies. August 22-28, 2022 is the week in which the S-DEIM reconstructions have minimum relative error.
	}
	\label{fig:recon_week34_anomaly}
\end{figure*}

In this appendix, we reproduce Figures~\ref{fig:recon_week21} and \ref{fig:recon_week34}, but this time we show the SST anomalies, instead of the difference between the reconstructions and the true SST. Figure~\ref{fig:recon_week21_anomaly} corresponds to the week of May 23-29, 2022 where the relative error of the reconstruction is maximum. On the other hand, Figure~\ref{fig:recon_week34_anomaly} corresponds to the week of August 22-28, 2022 where the relative error of the reconstruction is minimum.

\section{Voronoi-CNN Architecture}\label{app:CNN_Arch}
In this section, we detail the CNN architecture used in the Voronoi-CNN of Section \ref{Comparisons}. The implementation follows the publicly available code provided by \citeA{Fukami2021}. The network input is the Voronoi-preprocessed sparse observations together with a binary sensor mask. Thus, each input snapshot has two channels on the $360 \times 180$ spatial grid. The first channel is a Voronoi tessellation SST anomaly field, where each grid point is assigned the SST anomaly of its nearest sensor. The second channel is a binary sensor mask, with value one at the sensor locations and zero elsewhere. The output of the network is a single-channel SST anomaly reconstruction.

    The CNN consists of eight convolutional layers as shown in Table \ref{tab:CNNArch}. All convolutional layers use $7 \times 7$ kernels with padding equal to $3$, so that the spatial resolution is preserved throughout the network. The first layer maps the two-channel input to $48$ feature channels. This is followed by six additional convolutional layers that each map $48$ channels to $48$ channels. Each of these hidden layers is followed by a ReLU activation function. Finally, the last layer maps the $48$ feature channels to a single SST field and does not use a ReLU activation. No pooling or downsampling layers are used, so the network output remains on the original $ 360\times 180$ grid. This network has a total of $684,769$ trainable parameters. These parameters are tuned during the training process, using the Adam optimizer with an early stopping criterion and a threefold cross-validation~\cite{Prechelt1998}.

    \begin{table}[]
        \centering
        \begin{tabular}{c c c c c}
        \hline
        Layer & Input Channels & Output Channels & Filter Size & Activation \\
        \hline
             Conv 1 & 2 & 48 & $7 \times 7$ & ReLU \\
             Conv 2 & 48 & 48 & $7 \times 7$ & ReLU \\
             Conv 3 & 48 & 48 & $7 \times 7$ & ReLU \\
             Conv 4 & 48 & 48 & $7 \times 7$ & ReLU \\
             Conv 5 & 48 & 48 & $7 \times 7$ & ReLU \\
             Conv 6 & 48 & 48 & $7 \times 7$ & ReLU \\
             Conv 7 & 48 & 48 & $7 \times 7$ & ReLU \\
             Output & 48 & 1 & $7 \times 7$ & --- \\
             \hline
        \end{tabular}
        \caption{Voronoi-CNN architecture}
        \label{tab:CNNArch}
    \end{table}


\end{document}